\documentclass{elsart}

\usepackage{graphicx}
\usepackage{pxfonts}
\usepackage{lineno}

\usepackage{amssymb}
\usepackage{footnote}
\usepackage{multirow}
\usepackage{varwidth}
\usepackage{array}
\usepackage{epstopdf}
\usepackage{url}

\journal{}

\begin{document}

\thispagestyle{empty}
\begin{Large}
	\textbf{DEUTSCHES ELEKTRONEN-SYNCHROTRON}
	
	\textbf{\large{Ein Forschungszentrum der Helmholtz-Gemeinschaft}\\}
\end{Large}

DESY 19-047

March  2019

\begin{eqnarray}
\nonumber &&\cr \nonumber && \cr \nonumber &&\cr
\end{eqnarray}
\begin{eqnarray}
\nonumber
\end{eqnarray}
\begin{center}
	\begin{Large}
		\textbf{Wigner Rotation: Theory and Application to Practical Relativistic Engineering Problems}
	\end{Large}
	\begin{eqnarray}
	\nonumber &&\cr \nonumber && \cr
	\end{eqnarray}
	
	%
	
	Evgeny Saldin
	%
	\textsl{\\Deutsches Elektronen-Synchrotron DESY, Hamburg}
	\begin{eqnarray}
	\nonumber
	\end{eqnarray}
	\begin{eqnarray}
	\nonumber
	\end{eqnarray}
	ISSN 0418-9833
	\begin{eqnarray}
	\nonumber
	\end{eqnarray}
	\begin{large}
		\textbf{NOTKESTRASSE 85 - 22607 HAMBURG}
	\end{large}
\end{center}
\clearpage
\newpage

\begin{frontmatter}



\title{Wigner Rotation: Theory and Application to Practical Relativistic Engineering Problems}


\author[DESY]{Evgeni Saldin}
\address[DESY]{Deutsches Elektronen-Synchrotron (DESY), Hamburg, Germany}

\begin{abstract}
	
As known, a composition of noncollinear Lorentz boosts does not result in a different boost but in a Lorentz transformation involving a boost and  a spatial rotation, the Wigner rotation. 
We point out the very important role that Wigner rotation plays in the analysis and interpretation of experiments with ultrarelativistic, modulated electron beams in  XFELs. In previous papers we showed that when the evolution of the  modulation wavefront in the case of non-collinear electron beam motion is treated according to relativistic kinematics, the orientation of the wavefront in the ultrarelativistic asymptotic is always perpendicular to the electron beam velocity. This  wavefront readjusting after electron beam kicking is one concrete example of Wigner rotation.
Many experts who learned the theory of relativity using  textbooks will find this statement disturbing at first sight. In fact, the standard expression for the Wigner rotation leads to incorrect result for the wavefront rotation. The results for the Wigner rotation in the Lorentz lab frame obtained by  many experts on special relativity such as Moeller and Jackson, are incorrect. 
They overestimate the angle of the Wigner rotation by a factor $\gamma$ compared to its real value, and the direction of the rotation is also determined incorrectly. 	   In 1959, Bargman, Michel, and Telegdi (BMT) proposed a consistent relativistic theory for the dynamics of the spin as observed in the lab frame,  which was successfully tested in experiments.	
The BMT equation naturally involves the Wigner rotation as the purely kinematic addition to the Larmor rotation which, in turn, is a consequence of interaction of the intrinsic magnetic  moment with the external magnetic field. 	
It is commonly believed that  the BMT equation contains the standard (and incorrect) result for the Wigner rotation in the Lorentz lab frame. The existing textbooks then suggest that the experimental test of the BMT equation is a direct proof of validity for the standard expression for Wigner rotation. Here we  focus on the analysis of the reason why authors of textbooks obtained an incorrect expression for the Wigner rotation. 
We demonstrate that the notion that the standard (incorrect) result for the Wigner rotation as an integral part of the BMT equation in most texts is based, in turn, on an incorrect physical argument. The aim of the present paper is to analyze the complicated situation relating to the use of the theory of relativity and, in particular, Wigner rotation theory, in accelerator physics.

\end{abstract}

%
%

%
\end{frontmatter}



\section{ Introduction }

\subsection{Wigner rotation. Two practical applications}

\subsubsection{Evolution of a modulated electron beam in XFELs}

As known, a composition of noncollinear Lorentz boosts does not results in a different boost but in a Lorentz transformation involving a boost and  a spatial rotation, the Wigner rotation  \cite{WI, WI1,WI2}. The Wigner rotation effect plays an essential role  in the description of extended  relativistic objects. But up to 21 st century there were no macroscopic objects possessing relativistic velocities, and there was a general belief that only microscopic particles  in experiments can travel at velocities close to that of light. The 2010s saw a rapid development of new laser light sources in the X-ray wavelength range. An X-ray free electron laser (XFEL) is an example where improvements in accelerator technology makes it possible to develop ultrarelativistic macroscopic objects with an internal structure (modulated electron bunches), and the relativistic kinematics  plays an essential role in their description. 
Relativistic kinematics enters XFEL physics in a most fundamental way through the Wigner rotation of the modulation wavefront, which,  in ultrarelativistic approximation, is closely associated to the relativity of simultaneity. 

When the trajectories of particles calculated in the Lorentz reference frame (i.e. an inertial frame where Einstein synchronization procedure is used to assign values to the time coordinate) they must include  relativistic kinematics effects such as relativity of simultaneity. 
In the ultrarelativistic asymptote, the orientation of the modulation wavefront , i.e the orientation of the plane of simultaneity, is always perpendicular to the electron beam velocity when the evolution of the modulated electron beam is treated using Lorentz coordinates.

We should remark that Maxwell's equations are valid only in Lorentz reference frames.
Einstein's time order should obviously be applied and kept in consistent way  both in dynamics and electrodynamics. It is important at this point to emphasize that the theory of relativity dictates that a modulated electron beam in the ultrarelativistic asymptote has the same kinematics, in Lorentz coordinates, as a laser beam. According to Maxwell's equations, the wavefront of a laser beam is always orthogonal to the propagation direction. In other words, the ultrarelativistic limit of a  modulated electron beam is the massless particle limit, which is the same,  for instance, as the photon case.

The explanation of the effect of wavefront readjusting is actually based on the use of a Lorentz boosts (i.e. it is explainable in terms of relativistic kinematics). This explanation describes how the direction of a wavefront depends on  the velocity of the modulated electron beam relative to the lab frame. There are several cases where a modulation wavefront readjusting can occur in XFELs, mainly through the introduction of an angular trajectory kick. 
Suppose the beam velocity is perpendicular to the wavefront of the modulation upstream of the kicker magnet. As seen from the lab frame, the wavefront of the beam modulation rotates relative to  the Cartesian axes of the Lorentz  lab  frame when a modulated electron beam is accelerated in the kicker field. 
In the case  of an arbitrary electron beam velocity $\vec{v}$, expression for the Wigner rotation is given by \cite{Rit}

\begin{eqnarray}
	\vec{\delta \Phi} = 	\left(1 - \frac{1}{\gamma} \right)\frac{\vec{v}\times d\vec{v}}{v^2}	
	=	\left(1 - \frac{1}{\gamma} \right) \vec{\delta \theta} ~.
	\label{R}
\end{eqnarray}

where $\gamma$ is the Lorentz factor, $d\vec{v}$ is the vector of  small  velocity change due to acceleration, $\Phi$  is the Wigner rotation angle of the  wavefront, and $\theta$ is the orbital angle of the particle in the lab frame. From Eq. (\ref{R}) follows that in the ultra relativistic limit $\gamma \longrightarrow \infty$, the wavefront rotates exactly as the velocity vector $\vec{v}$.

The majority of authors  follow the incorrect expression for Wigner rotation obtained by  many experts on special relativity such as Moeller \cite{M}, Jackson \cite{JACK}, and  Sard \cite{SA}, which is given by  $\vec{\delta \Phi} =  (1 - \gamma) \vec{v}\times d\vec{v}/v^2 =	
(1 - \gamma) \vec{\delta \theta}$.
It should be note that, in \cite{M,JACK,SA} authors specified quite clearly   that this expression is valid in the Lorentz lab frame. Clearly, it differs compared with Eq. (\ref{R})  both in sign and in magnitude. 
In review \cite{MAL} it is shown that the correct result Eq. (\ref{R})  was obtained in the works of several authors, which were published more than half century ago but remained unnoticed against the background of numerous incorrect works.


From Eq. (\ref{R}) follows that,  in the ultrarelativistic asymptote, we have a limit where our modulated electron beam approaches a beam of massless particles. In contrast, 
according to  the incorrect (but conventional) expression reported before, the modulation wavefront rotates in opposite direction  and $\Phi = (1- \gamma\theta) \to -\infty$ in the limit $\gamma \longrightarrow \infty$. This contradicts both  common sense and  Maxwell's electrodynamics.

Here we will give a simple proof of the conflict between conventional expression for the Wigner rotation in the  Lorentz  lab frame and Maxwell's electrodynamics.

Under the Maxwell's electrodynamics, the fields of a modulated electron beam moving with a constant velocity exhibit an interesting behavior when the velocity of charges approaches that of light: namely, in the space-time domain they resemble more and more close of a laser beam. In fact,
for a rapidly moving  modulated electron beam we have  nearly equal transverse and mutually perpendicular electric and magnetic fields: in the limit $v \to c$ they become indistinguishable  from the fields of a laser beam, and according to Maxwell's equations, the wavefront of the laser beam is always perpendicular to the propagation direction
\footnote{How to find a polarization of a virtual radiation beam is an interesting question. Within the deep asymptotic region when the transverse size of the modulated electron beam $\sigma \ll \lambdabar\gamma$ the Ginzburg-Frank formula can be applied. Here $\lambda$ is the modulation wavelength.  In this asymptotic region radiation can be considered as virtual radiation from a filament electron beam (with no transverse dimensions).  However, in XFEL practice we only deal with the deep asymptotic region where $\sigma \gg  \lambdabar\gamma$. Then, it can be seen that the field distribution in the space-time domain is essentially a convolution in the space domain between the transverse charge distribution of the electron beam and the field spread function described by the Ginzburg-Frank formula. Assuming a Gaussian (azimuthally-symmetric) transverse density distribution of the electron beam we obtain the
radially polarized virtual radiation beam.}.
This is indeed the case for virtual laser-like radiation beam  in the region upstream the kicker. 

Let us now consider the effect of the kick on the electron modulation wavefront. If we rely on the conventional expression for Wigner rotation, the kick results in a difference between the directions of  electron motion and the normal to the modulation wavefront, i.e.  in a  tilt of the modulation wavefront.

This is already a conflict result, because we now conclude that, according to the conventional expression for the Wigner rotation, the direction of propagation after the kick is not perpendicular to the radiation beam wavefront.  
In other words, the radiation beam motion and the radiation wavefront normal have different directions.  The virtual radiation beam (which is indistinguishable from a real radiation beam in the ultrarelativistic asymptote) propagates in the kicked direction with a wavefront tilt.
This is what we would get for the case when our analysis is based on the conventional expression for the Wigner rotation in the  Lorentz lab frame, and is obviously absurd from the viewpoint of  Maxwell's electrodynamics. We conclude that the conventional expression for the Wigner rotation is incorrect.

An analysis of the reason why authors of famous  textbooks obtained an incorrect expression for the Wigner rotation in the Lorentz lab frame is the focus of Ritus paper \cite{Rit}. As shown in \cite{Rit}, the  mistake of acknowledged experts on the theory of relativity is not computational, but conceptual in nature. The explanation of the effect of modulation wavefront readjusting is  described by correct formula Eq. (\ref{R}).

Above we pointed out that if the velocity of our modulated electron beam is close to velocity of light, Lorentz transformations work out in such a way that the rotation angle of the modulation wavefront coincides with the angle of rotation of the velocity. 
Experiment shows that this prediction is, in fact, true.
In XFEL simulations it is generally accepted that coherent radiation from the undulator placed after the kicker is emitted ( in accordance with Maxwell's electrodynamics) along the normal to the modulation wavefront. Therefore, when the tilt angle exceeds the divergence of the output radiation, emission in the kicked direction is suppressed.  
The  experiment at the LCLS \cite{NUHN}  apparently demonstrated that after a modulated electron beam is kicked on a large angle compared to the divergence of the XFEL radiation, the modulation wavefront is readjusted along the new direction of motion of the kicked beam, see Fig. 14 in \cite{NUHN}.

\subsubsection{Spin evolution in storage rings}

Relativistic effects start to be important when velocities of objects get closer to the speed of light. Usually, only  elementary particle velocities may be a substantial fraction of the speed of light. Consider a particle with its own angular momentum (spin).
In 1959, a paper by Bargmann, Michel, and Telegdi \cite{BMT} was published, which dealt   
with the motion of elementary charged spinning particles with an anomalous magnetic moment in electromagnetic field. The BMT equation, describing such motion, is manifestly covariant and can be used in any inertial frame. It is the law of motion of the four-spin for a particle in a uniform electric and magnetic fields. The extremely precise measurements of the magnetic-moment anomaly of the electron made on highly relativistic electrons are based on the BMT equation.
The anomalous magnetic moment can be calculated by use of quantum electrodynamics. 
The theoretical result agrees with experiments to within a very high accuracy. This can be regarded as a direct test of BMT equation.

The formalism of the Wigner rotation has been applied to physical problems involving the  dynamics equation describing a magnetic moment in an electromagnetic field. 
It is known that Wigner rotation represents a kinematics effect, due to the fact that the successive Lorentz transformations with non-collinear relative velocities are accompanied by an additional spatial rotation of coordinate axes of corresponding reference frames.
The Wigner rotation is a purely relativistic kinematics effect but it influences the dynamics of the spin as observed in the lab frame. 	The BMT equation naturally involves the Wigner rotation, since this is  the purely kinematic addition to the Larmor rotation which, in turn, is a consequence of interaction of the intrinsic magnetic moment with the external magnetic field. 	

Today one is told that the standard expression for Wigner rotation in the Lorentz lab frame, entering textbooks on relativistic mechanics, is indeed included in the BMT equation.
We should remark that the results in the Bargmann-Michel-Telegdi  paper \cite{BMT} were obtained by the method of semi-classical approximation of the Dirac equation. The Wigner rotation was not considered in \cite{BMT} at all, because the Dirac equation allow obtaining the solution for the total particle's spin motion without an explicit splitting it into the Larmor and Wigner parts. Apparently the success of the BMT equation made people less interested in probing the details of the derivation of the expression for Wigner rotation. The existing textbooks suggest that the experimental test of the BMT equation is a direct test of what we consider the incorrect expression for Wigner rotation in the Lorentz lab frame. We claim  that the inclusion of this incorrect expression  as an integral part of the BMT equation in most texts is based on  an incorrect physical argument. In this paper we will investigate in detail the reason why this is the case.

\subsection{Modulation wavefront evolution and conventional particle tracking}

The use of the theory of relativity in the problem of the rotation of the modulation wavefront of an electron beam in an XFEL is  complicated conceptually. 
The usual analysis of relativistic particle motion in a constant magnetic field  performed in relativistic engineering looks precisely the same as in non-relativistic Newtonian dynamics and kinematics. The solution of the dynamics problem in the lab frame makes no reference  to Lorentz transformations, and the trajectory of the particle does not include  relativistic kinematics effects. 

As well-known, according to conventional particle tracking in the lab frame, after the electron beam is kicked by a weak dipole magnet (kicker) there is a change in the trajectory of the electron beam, while the orientation of the modulated wavefront remains as before. According to conventional particle tracking, within the lab frame there is no Wigner rotation. This, as we already mentioned, contradicts  the Maxwell's electrodynamics. 
We cannot take old kinematics for mechanics and Einstein's kinematics for electrodynamics.
Therefore, something in accelerator physics is  fundamentally  wrong. If one wants to use the usual Maxwell's equations, only solution of the dynamics equations in covariant form gives the correct coupling between Maxwell's equations and particle trajectories in the lab frame.

Many accelerator physicists, who have not had further training in theoretical physics, find that even the fact that in literature  there are  two  expressions for the Wigner rotation in the lab frame which differ both in sign and in magnitude
is a barrier in the further understanding of the situation relating to the use of the theory of relativity in the problem of  rotation of the modulation wavefront of an electron beam in an XFEL. In no way to the two quantitatively different expressions for the Wigner rotation lead to the same result in the same frame of reference.  In this paper, therefore, we consider the correct Wigner rotation theory in detail. We will see that one of the large stumbling block in the way of acceptance of the correct Wigner rotation theory  is a myth about experimental test of the incorrect result for the Wigner rotation. 
Our aim is to point out that, after removing this "stumbling block" the path is open to viewing the practical applications of the Wigner rotation theory in an XFEL physics. It is hoped that our publication can become a useful and physically constructive supplement to our previous publications \cite{OURS5,OURS6}, where we analyzed the complicated situation relating to the use of the theory of relativity in accelerator physics.

\section{A non-covariant approach to relativistic particle dynamics}

Before treating the case of a spinning charged particle we study the motion of a charged point particle in an electromagnetic fields.
In this section we discuss first the accelerated motion in a constant magnetic field - the subject of  the conventional particle tracking in accelerator physics.

The accelerated motion is described by a covariant equation of  motion for a relativistic charged particle under the action of the four-force in the Lorentz lab frame. The trajectory of a particle $\vec{x}_{cov}(t)$ is viewed from the  Lorentz lab  frame as a result of successive infinitesimal Lorentz transformations. Under the Einstein's synchronization convention the lab frame time $t$ in the equation of motion cannot be independent from the space variables. This is because Lorentz transformations lead to a mixture of positions and time,  and the relativistic kinematics effects are considered to be a manifestation of the relativity of simultaneity.

Let us consider the conventional particle tracking approach.
This solution of the dynamics problem in the lab frame makes no reference  to Lorentz transformations. It is generally accepted that in order to describe the dynamics of relativistic particles in the lab reference frame, one only needs to take into account the relativistic dependence of the particles momenta on the velocity. In other words, the treatment of relativistic particle dynamics involves only a corrected Newton's second law. Conventional particle tracking treats the space-time continuum  in a non-relativistic format, as  a (3+1) manifold.
In other words, in this approach, introducing as only modification  to the classical case the relativistic mass, time differ from space. In fact, we have no mixture of positions and time. 
To quote Feynman, Leiton and Sands \cite{F}: "Newton's second law, $d(m\vec{v})/dt = \vec{f}$, was stated with the tacit assumption that $m$ is a constant, but we now know that this is not true, and the mass of a body increases with velocity. (...) For those who want to learn just enough about it so they can solve problems, that is all there is to the theory of relativity - it just changes Newton's laws by introducing a correction factor to the mass."

Most of the interesting phenomena in which charges move under the action of electromagnetic fields occur in very complicated situations. But here we just want to discuss the simple problem of the accelerated motion of particles in a constant  magnetic field. 
According to the non-covariant treatment, the magnetic field  is only capable of altering the direction of motion, but not the speed (i.e. mass) of a particle. 
This study of relativistic particle motion in a constant magnetic field, usual for accelerator engineering,  looks precisely the same as in nonrelativistic Newtonian dynamics and kinematics. 
The trajectory of a particle $\vec{x}(t)$, which follows from the solution of the corrected Newton's second law, does not include  relativistic kinematics effects.

\subsubsection{Difference between covariant and non-covariant trajectories}

This point was expressed by Friedman \cite{Frid}: "Within any single inertial frame, things looks precisely the same as in Newtonian kinematics: there is an enduring Euclidean three-space, a global (i.e. absolute) time $t$, and law of motion. But different inertial frames are related to one another in a non-Newtonian fashion." Similar explanations can be found in various other advanced textbooks , see e.g. \cite{RCM}. 

Conventional particle tracking states that after the electron beam is kicked by a weak dipole magnet (kicker) there is trajectory change,  while the orientation of the modulated wavefront remains as before. In other words, the kick results in a difference between the directions of the electron motion and the normal to the modulation wavefront (i.e. in a wavefront tilt). According to conventional particle tracking, within the lab frame, the particles motion follows the corrected Newton equations, and there is no Wigner rotation. 

In order to obtain relativistic kinematics effects, and in contrast to conventional particle tracking, one actually needs to solve the dynamics equation in covariant form by using the coordinate independent proper time as evolution parameter. In the case of Einstein's synchronization convention the covariant trajectory $\vec{x}(t)_{cov}$ is viewed from the Lorentz lab frame as  a result of successive Lorentz transformations.   
We know that, in the ultrarelativistic asymptote, the orientation of the modulation wavefront is always perpendicular to the electron beam velocity when the evolution of the modulated electron beam is treated using Lorentz coordinates.


Even  for a single particle we are able to demonstrate the difference between conventional and covariant particle tracking results. In fact, we use Einstein's rule for adding velocities to track the particle motion in a covariant way. But in the conventional particle tracking the velocity summation is curried out differently. In accelerator physics the dynamical evolution in the lab frame is based on the usual Galileo (vectorial) rule which is in agreement with velocity summations of Newtonian mechanics. In particular, think of the algorithm that one actually uses while updating the velocity from one moment in time to the next in conventional particle tracking: one just uses the Galilean law of addition of velocities, not Einstein's one.
In contrast to this, in the case of covariant particle tracking relativistic kinematics effects arise and the covariant trajectory $\vec{x}_{cov}(t)$ is viewed from the lab frame as a result of successive Lorentz transformations.

So we must conclude that for the accelerated motion in a constant magnetic field the covariant trajectory of the particle $\vec{x}(t)_{cov}$ and result from conventional (non-covariant) particle tracking $\vec{x}(t)$ differ from each other.

\subsubsection{Mistake in commonly used method of coupling fields and particles}

It is generally believed  that the electrodynamics problem can be treated within the same "single inertial frame"  description without reference to Lorentz transformations (see the standard textbooks, e.g. \cite{JACK}). In all standard derivations it is assumed  that usual Maxwell's equations and corrected Newton's second law can explain all experiments that are performed in a single inertial frame, for instance the lab reference frame. In particular, in order to evaluate  radiation fields arising from a collection of sources  we only need to specify   velocities  and the positions of the charged particles  as a function of the lab frame time $t$. In its turn, the relativistic motion of these particles in the lab frame is described by the corrected Newton's second law. This coupling of Maxwell's equations and corrected Newton's equation  is commonly accepted as useful method  in accelerator physics and, in particular, in analytical and numerical calculations of  radiation properties. Such approach to relativistic dynamics and electrodynamics usually forces the accelerator physicist to believe that the design of particle accelerators possible without detailed knowledge of the theory of relativity.

However, there is a common mistake made in accelerator physics connected with the difference between $\vec{x}(t)$ and $\vec{x}_{cov}(t)$ trajectories. Let us look at this difference from the point of view of electrodynamics of relativistically moving charges. To evaluate fields arising from external sources we need to know their velocity and positions as a function of the lab frame time $t$. Suppose one wants to calculate properties of  radiation. Given our previous discussion the question arises, whether one should solve the usual Maxwell's equations in the lab frame with current and charge density created by particle moving along non-covariant trajectories like $\vec{x}(t)$. We claim that the answer to this question is negative.
In our previous publications \cite{OURS5,OURS6} we argued that
this algorithm for solving usual Maxwell's equations in the lab frame, which is considered in all standard treatments as relativistically correct, is at odds with the principle of relativity. 
This essential point has never received attention in the physical community. Only the solution of the dynamics equations in covariant form gives the correct coupling between the usual Maxwell's equations and particle trajectories in the lab frame.

\subsection{Misconception regarding the "single frame" approach in accelerator physics}

\subsubsection{Maxwell's electrodynamics and absence of relativistic kinematics effects}

We found that the usual integration of the  corrected Newton equation gives   particle trajectories which looks precisely the same as in Newton mechanics. The trajectories does not include such  relativistic kinematics effect as Wigner rotation and the Galilean vectorial law of addition of velocities is actually used. 
To quote Feynman, Leiton and Sands \cite{F}: "How shell we change Newton's laws so that they will remain unchanged by the Lorentz transformations? If this goal is set, we then have to rewrite Newton's equations in such a way that the conditions we have imposed are satisfied. As it turned out, the only requirement is that the mass $m$ in Newton's equations must be replaced by the form shown in Eq.(15.1) $m = m_0/\sqrt{1 - v^2/c^2}$. When this change will made, Newton's laws and the laws of Maxwell's electrodynamics will harmonize". 

It is impossible to agree with this textbook statement. 
The central principle of special relativity is the Lorentz covariance of all the fundamental laws of physics. Let us discuss the consequences of Lorentz transformations. In particular we discuss the important problem of the addition of velocities in relativity. 
Suppose that in the  case of accelerated motion one introduces an infinite sequence of co-moving frames. At each instant, the rest frame is a Lorentz frame centered on the particle and moving  with it.  Suppose that in inertial frame where particle is at rest at a given time, the traveler was observing light itself. In other words measured speed of light $v = c$, and yet the frame is moving relative the lab frame. How will it look to the observer in the lab frame? According to Einstein's law of addition of velocity the answer will be $c$.  Maxwell's equations remain in the same form when Lorentz transformations are applied to them, but Lorentz transformations give rise to non-Galilean transformation rules for velocities, and therefore the theory of relativity shows that, if  Maxwell's equations is to be valid in the lab frame, the trajectories of the particles must include relativistic kinematics effects. In other words, Maxwell,s equations can be applied in the lab frame only in the case when particle trajectories are viewed, from the lab frame as the result of successive infinitesimal Lorentz transformations.

The absence of relativistic kinematics effects is the prediction of conventional non-covariant theory and is obviously absurd from the viewpoint of  Maxwell's electrodynamics. Therefore, something is fundamentally, powerfully, and absolutely  wrong in coupling fields and particles within a "single inertial frame".

\subsubsection{Choice of coordinates system in an inertial frame}

We will begin introducing   space and time coordinates of events in the lab  reference frame, as well as space and time coordinates of the same events in the comoving reference frame (one of the infinite sequence described above) where particle is at rest by definition.
This will be firm basis for our demonstration that the algorithm of solving an electrodynamics problem within the standard, non-covariant theory  contradicts the principle of relativity.

Each physical phenomenon occurs in space and time. A concrete method for representing space and time is a frame of reference. One-and-the same space and time can be represented by various coordinate-time grids, i.e., by various frames of references. Even the simplest space-time coordinate systems require carefully description. 
Clocks reveal the motion of a particle through the coordinate-time grid. The general approach to the determination of the motion of a particle is the following: at any instant a particle has a well-defined velocity $\vec{v}$  as measured in a laboratory frame of reference. How is a velocity of a particle found? The velocity  is determined once the coordinates in the  lab frame are chosen, and is then  measured at appropriate time intervals along the particle's trajectory. But how to measure a time interval between events occurring at different points in space? In order to do so, and  hence measure  the velocity of a particle within a single inertial lab frame, one first has to synchronize distant clocks.  The concept of synchronization is a key concept in the understanding of special relativity. It is possible to think of various methods to synchronize distant clocks. The choice of a convention on clock synchronization is nothing more than a definite choice of coordinates system in an inertial frame of reference of the Minkowski space-time \cite{LAK}.

The space-time continuum can be described in arbitrary coordinates. By changing these arbitrary coordinates, the geometry of the four-dimensional space-time obviously does not change, and in the special theory of relativity we are not limited in any way in the choice of a coordinates system. The space coordinates $x^1, x^2, x^3$ can be any quantities defining the position of particles in space, and the time coordinate $t$ can be defined by an arbitrary running clock.
Relying on the geometric structure of Minkowski space-time, one defines the class of inertial frames and adopts a Lorentz frame with orthonormal basis vectors. Within the chosen Lorentz frame, Einstein's synchronization procedure of distant clocks (which based on the constancy of the speed of light in all inertial framers) and Cartesian space coordinates are enforced: 
covariant particle tracking is based on the use Lorentz coordinates. 

\subsubsection{What is meant when it is said that an inertial frame is "single" }

Clarification of the true content of the non-covariant theory can be found in various advanced textbooks. To quote e.g. Ferrarese and Bini \cite{RCM}: "... within a single inertial frame, the time is an absolute quantity in special relativity also. As a consequence, \textit{if no more than one frame is involved}, one would not expect differences between classical and relativistic kinematics. But in the relativistic context there are differences in the transformation laws of the various relative quantities (of kinematics or dynamics), when passing from one reference frame to another."

We see that authors give a special role to concept of a "single inertial frame". 
The name "single inertial frame" tends to suggest that a distinctive trait of  non-covariant theory is the absence of relativistic kinematics in the description of particle motion.  In fact, kinematics is a comparative study. It requires two relativistic observers and two coordinate systems. We now examine the logical content of the concept of a "single inertial frame".

Disproof of the conventional non-covariant (single frame) theory is related to a clear statement in  textbook \cite{RCM} " \textit{if no more than one frame is involved}, one would not expect differences between classical and relativistic kinematics." 
If a traveler in a co moving frame, similar to an observer in the lab frame, introduces a definite coordinate-time grid, there is always a definite transformation between these two four-dimensional coordinate systems. Thus, particle trajectories are always viewed from the lab frame as a result of successive transformations, and the form of these transformations depends on the choice of coordinate systems in the comoving and the lab frame. 

One might well wonder why it is necessary to discuss how different inertial frames are related to one another. The point is that all natural phenomena follow the principle of relativity, which  is a restrictive principle: it says that the laws of nature are the same (or take the same form) in all inertial frames. In agreement with this principle, usual Maxwell's equations can always be exploited in any inertial frame where electromagnetic sources are at rest using Einstein synchronization procedure in the rest frame of the source. The fact that one can deduce electromagnetic field equations for arbitrary moving sources  by studying the form taken by Maxwell's equations under the transformation between rest frame of the source and the frame where the source is moving is a practical application of the principle of relativity. Since we require two coordinate systems, the question now arises how to assign a time coordinate to the lab frame. 

Coordinates serve the purpose of labeling events in an unambiguous way, and this can be done in infinitely many different ways. The choice made in different cases is only a matter of convenience. However, we are better off using Lorentz coordinates when we want to solve the electrodynamics problem in the lab frame based on Maxwell's equations in their usual form. In fact, the use of other coordinate systems also implies the use of much more complicated electromagnetic field equations.    
In other words, the principle of relativity dictates that Maxwell's equations can be applied in the lab frame only in the case when  Lorentz coordinates are assigned and particle trajectories are viewing from the lab frame as a result of successive infinitesimal Lorentz transformations between the lab and comoving inertial frames.

A non-covariant (3+1) approach to relativistic particle dynamics has been used in particle tracking calculations for about seventy years. However, the type of clock synchronization which provides the time coordinate $t$ in the corrected Newton's equation has never been discussed in literature. It is clear that without an answer to the question about the method of synchronization used, not only the concept of velocity, but also the dynamics law has no physical meaning. 
A  non-covariant (3+1) approach to relativistic particle dynamics is forcefully based on a definite synchronization assumption but this is actually a hidden assumption.  
According to conventional particle tracking, the dynamical evolution in the lab frame is based on the use of the lab frame time $t$ as an independent variable, independent in the sense that $t$ is not related to the spatial variables. Such  approach to relativistic particle dynamics   is actually based on the use of a not standard (not Einstein) clock synchronization assumption in the lab frame.

\subsubsection{What is meant when it is said that time is "absolute"}

We should make one further remark about the textbook statement  \cite{RCM}: "... within a single inertial frame, the time is an absolute quantity in special relativity also." 
What does "absolute" time mean? It means that simultaneity is absolute and there is no mixture of positions and time when particles change their velocities in the lab frame.
In fact, the usual for accelerator engineering study of relativistic particle motion in a constant magnetic field looks precisely the same as in nonrelativistic Newtonian mechanics and the trajectories of the electrons does not include relativistic kinematics effects. According to textbooks, this is no problem. If no more than one frame is involved, one does not need to use (and does not need to know) the theory of relativity. 
Only when one  passes from one reference frame to another the relativistic context is important.
Conventional particle tracking in a constant magnetic field is actually based on  classical Newton mechanics. It is generally believed that the electrodynamics problem, similar to conventional particle tracking, can be treated within a description involving a single inertial frame and one should solve the usual Maxwell's equations in the lab frame with current and charge density created by particles moving along the non-covariant (single frame) trajectories.

This is misconception.  The situation when only one frame is involved and the relativistic context is unimportant cannot be realized. The lab observer may argue, "I don't care about other frames." Perhaps the lab observer doesn't, but nature knows that, according to the principle of relativity,  Maxwell's equations are always valid in the Lorentz comoving frame. Electrodynamics equations can be written down in the lab frame only when a space-time coordinate system has been specified.
An observer in the lab frame has only one freedom. This is the choice  of a coordinate system (i.e. the choice of clock synchronization convention) in the lab frame. After this,  the theory of relativity states that the electrodynamics equations in the lab frame are the result of transformation of Maxwell's equations from the Lorentz comoving frame to the lab frame. 

\subsection{Change to a different four-dimensional coordinate systems}

We must emphasize that
in special relativity there are two choices of clock synchronization convention useful to consider:

(a) Einstein's convention,  leading to the Lorentz  transformations between frames

(b) Absolute time convention, leading to the Galilean transformations between frames

Absolute time (or simultaneity) can be introduced in special relativity without affecting neither the logical structure, no the  (convention-independent) predictions of the theory. Actually, it is just a simple effect related with a particular parametrization. In the theory of relativity this choice may seem quite unusual, but it is usually most convenient when one wants to connects to laboratory reality.  As matter of fact it is this hidden synchronization convention that is used, practically, in accelerator physics.

 When a time coordinate is assigned in the lab frame, non-covariant particle trajectories can be experimentally interpreted by a laboratory observer. Due to the particular choice of synchronization convention, relativistic kinematics effects such as relativity of simultaneity and Einstein's edition of velocities do not exist in the lab frame. 
Particle tracking calculations usually become much simpler if the particle beam evolution is treated in terms of absolute time (or simultaneity). This time synchronization convention is self-evident and this is the reason why it is never discussed in relativistic engineering. In non covariant particle tracking, time differs from space and a particle trajectory in a constant magnetic field can be seen from the lab frame as the result of successive Galilean boosts that track the accelerated motion.

\subsection{Myth about the incorrectness of Galilean transformations}

The use of Galilean transformations within the theory of relativity requires some special discussion. Many physicists still tend to think of Galilean transformations as old, incorrect transformations between spatial coordinates and time. 
A widespread argument used to support the incorrectness of Galilean transformations is that they do not preserve the form-invariance of Maxwell's equations under a change of inertial frame. This idea is a part of the material in well-known books and monographs. To quote e.g.  Bohm \cite{Bohm} "... the Galilean law of addition of velocities implies that the speed of light should vary with the speed of the observing equipment. Since this predicted variation is contrary to the fact, the Galilean transformations evidently cannot be the correct one.".  Similar statements can also be found in recently published  textbooks. To quote e.g.  Cristodoulides \cite{Cr} "The fact that Galilean transformation does not leave Maxwell's equations has already been mentioned [...] On the other hand, experiments show that the speed of light in vacuum is independent of the source or observer.".

Authors of textbooks are mistaken in their belief about the incorrectness of Galilean transformations. 
The special theory of relativity is the theory of four-dimensional space-time with pseudo-Euclidean  geometry. From this viewpoint, the principle of relativity is a simple consequence of the space-time geometry, and the space-time continuum can be described in arbitrary coordinates. Therefore, contrary to the view presented in most textbooks,  Galilean transformations are actually compatible with the principle of relativity although, of course, they alter the form of Maxwell's equations.

The mathematical argument that in the process of transition to arbitrary coordinates the geometry of the space-time does not change, is considered in textbooks as erroneous.
To quote L. Landau and E. Lifshitz \cite{LL}:  "This formula is called the Galileo transformation. It is easily to verify that this transformation, as was to be expected, does not satisfy the requirements of the theory of relativity; it does not leave the interval between events invariant.".  
This statement indicates that the authors of textbook do not understand that the space-time continuum can be described  equally well from the point of view of any coordinate system, whose choice  cannot change  geometry. In pseudo-Euclidean  geometry
the interval between events is an invariant in arbitrary coordinates.

We emphasize the great freedom one has in the choice of a Minkovski space-time coordinatization.
In order to develop space-time geometry, it is necessary to introduce a metric or a measure $ds$ of space-time intervals. The type of measure determines the nature of the geometry.  
The evolution of a particle is represented by a curve in space-time, called world-line. If $ds$ is the infinitesimal displacement along a particle world-line, then

\begin{eqnarray}
	&& ds^2 =  c^2 dT^2 - dX^2 - dY^2 - dZ^2~ ,\label{MM1}
\end{eqnarray}
where we have selected a special type of coordinate system (a Lorentz coordinate system),  defined by the requirement that Eq. (\ref{MM1}) holds.

To simplify our writing we will use, instead of variables $T, X, Y, Z$,  variables $X^{0} = cT, ~ X^{1} = X,~ X^{2} = Y,~ X^{3} = Z$. Then, by adopting the tensor notation, Eq. (\ref{MM1}) becomes $ds^2 = \eta_{ij}dX^{i}dX^{j}$, where Einstein summation is understood. Here $\eta_{ij}$ are the Cartesian components of the metric tensor and by definition, in any Lorentz system, they are given by $\eta_{ij} = \mathrm{diag}[1,-1,-1,-1]$, which is the metric canonical, diagonal form.

The components of the metric tensor in another coordinate system $x^i$ can be determined by performing the transformation from the Lorentz coordinates  $X^{i}$ to the arbitrary variables $x^{j}$, which are fixed as $X^{i} = f^{i}(x^{j})$. One then obtains

\begin{eqnarray}
	&& ds^2 = \eta_{ij}dX^{i}dX^{j} = \eta_{ij}\frac{\partial X^{i}}{\partial x^{k}}\frac{\partial X^{j}}{\partial x^{m}}dx^{k}dx^{m} = g_{km}dx^{k}dx^{m} ~ ,\label{MM3}
\end{eqnarray}

This expression represents the general form of the pseudo-Euclidean metric.
In the space-time geometric approach, special relativity is understood as a theory of  four-dimensional space-time with pseudo-Euclidean geometry. In this formulation of the theory of relativity the space-time continuum can be described  equally well from the point of view of any coordinate system, which cannot possibly change $ds$.

Here we only wish to show how naturally absolute time   can be introduced in  special relativity.
How can we find the components of the metric tensor in the case of the absolute time coordinatization? We begin with the Minkowski metric as the true measure of space-time intervals for an inertial observer $S'$ with coordinates $(t',x')$. Here we neglect the two perpendicular space components that do not enter in our reasoning. We transform coordinates $(t,x)$ that would be coordinates of an inertial observer $S$ moving with velocity $-v$ with respect to the observer $S'$, using a Galilean transformation: we substitute $x' =x-vt$, while leaving time unchanged $t' = t$ into the Minkowski metric $ds^2 = c^2 dt'^2 - d x'^2$ to obtain $ds^2 = c^2(1-v^2/c^2)dt^2 + 2vdxdt - dx^2$. Inspecting this equation we can find the components of the metric tensor $g_{\mu\nu}$ in the coordinate system $(ct,x)$ of $S$. We obtain $g_{00} = 1-v^2/c^2$, $g_{01} = v/c$, $ g_{11} = - 1$. Note that the metric  is not diagonal, since, $g_{01} \neq 0$, and this implies that time is not orthogonal to space.

The speed of light emitted by a  moving source measured in the lab frame $(t, x)$ depends on the relative velocity of source and observer, in our example $v$. In other words, the speed of light is compatible with the Galilean law of addition of velocities. The reason why it is different from the electrodynamics constant $c$
is due to the fact that the clocks are synchronized following the absolute time convention, which is fixed because $(t, x)$ is related to $(t',x')$ via a Galilean transformation. 
Note that from what we just discussed follows the statement that the difference between the speed of light and the electrodynamics constant $c$ is convention-dependent and has no direct physical meaning.

\subsection{Myth about the non-relativistic limit of the Lorentz transformations}

It is generally believed that a Lorentz transformation reduces to a Galilean transformation in the non-relativistic limit. To quote e.g. French \cite{FR}:  "The reduction of $t'= \gamma(t - vx/c^2)$ to Galilean relation $t' = t$ requires $x \ll ct$ as well as $v/c \ll 1$". 
Similar statements can also be found in recently published textbooks. To quote e.g. Henriksen \cite{HE}: "... the Lorentz transformations between any two inertial reference frames must replace the familiar Galilean transformations. It is only true at high velocities since the Lorentz transformations reduce to the Galilean ones to first order in $v/c$."    

We state that this typical textbook statement is incorrect and misleading. As discussed, kinematics is a comparative study which requires two coordinate systems, and one needs to assign time coordinates to the two systems. Different types of clock synchronization provide different time coordinates.
The convention on the clock synchronization amounts to nothing more than a definite choice of the coordinate system in an inertial frame of reference in Minkowski space-time. Pragmatic arguments for choosing one coordinate system over another may therefore lead to different choices in different situations. Usually, in relativistic engineering, we have a choice between absolute time coordinate and Lorentz time coordinate. The space-time continuum can be described equally well in both coordinate systems. This means  that  for arbitrary particle speed,  the Galilean coordinate transformations well characterize a change in the reference frame from the lab inertial observer to a comoving inertial observer in the context of the theory of relativity. Let us consider the non relativistic limit. The Lorentz transformation, for $v/c$ so small that $v^2/c^2$ is neglected can be written as $x' = x - vt$, $t' = t - xv/c^2$. This infinitesimal Lorentz transformation differs from the infinitesimal Galilean transformation $x' = x - vt$, $t' = t$. The difference is in the term $xv/c^2$ in the Lorentz transformation for time, which is a first order term.

We want next to use the Lorentz transformations for a special case - to find relativistic kinematics effects when the particle is accelerating from rest.
The appearance of a relativistic effects does not depends on the use a large relative speed of the two reference frames. Suppose that the particle is at rest in an inertial frame for an instant. At this instant, one picks a Lorentz coordinates system. Then, an instant later, the particle velocity changes of an infinitesimal value $dv$ along the $x$-axis.  The Lorentz transformation describing this change, for $dv/c$ so small that $dv^2/c^2$ is neglected, is described by $x' = x - ct(dv/c), ~ ct' = ct - x(dv/c)$.   The relativity of simultaneity is, then, the only relativistic effect that appears in the first order in $dv/c$.   To obtain a transformation valid for a finite relative speed between two reference frames, we must consider $n$ successive infinitesimal transformations, and then take the limit $n \longrightarrow \infty$, $dv/c \longrightarrow 0$, $ndv/c  \longrightarrow v/c$.
Consider first the case in which $v/c$ is fairly small, so that we neglect $v^3/c^3$, but not $v^2/c^2$. This case yields effects of the second order, which need to be considered in addition to the relativity of simultaneity, which appears already in the first order. Also time dilation and length contraction appear in the second order $v^2/c^2$, while the relativistic correction in the composition of velocities only appears in the order $v^3/c^3$ and higher. If the increments of the velocity are not all in the same direction, the transformation matrices do not commute, and this originates a Wigner rotation, which also appears in the order $v^2/c^2$.

We only wish to emphasize here the following point. 
An infinitesimal Lorentz transformation  differs from Galilean transformation only by the inclusion of the relativity of simultaneity, which is the only relativistic effect that appears in the first order in $v/c$.  All other higher order effects, that are Wigner rotation, Lorentz-Fitzgerald contraction, time dilation, and relativistic correction in the law of composition of velocities,  can be derived mathematically, by iterating this infinitesimal transformation.

The main difference between the Lorentz coordinatization and the absolute time coordinatization 
is that the transformation laws connecting  coordinates and times between relatively moving systems are different. It is impossible to agree with the following textbook statement \cite{FR}: "There is  reduction of $t'= \gamma(t - vx/c^2)$ to Galilean relation $t' = t$ in the non-relativistic limit." This would mean that in the non-relativistic limit  infinitesimal  Lorentz transformations are identical to infinitesimal Galilean transformations. 
This statement is absurd conclusion from a mathematical standpoint, and indicates that the authors of textbooks do not understand that the essence of Lorentz (or Galilean) transformations consists in their infinitesimal form: relativistic kinematics effects cannot be found by the mathematical procedure of iterating the infinitesimal Galilean transformations.

\subsection{Myth about the constancy of the speed of light}

 It is generally believed that "experiments show that the speed of light in vacuum is independent of the source or observer."\cite{Cr}. This statement presented in most textbooks and is obviously incorrect.
The constancy of the speed of light is related to the choice of synchronization convention, and cannot be subject to experimental tests.

In fact, in order to measure the one-way speed of light one has first to synchronize the infinity of clocks assumed attached to every position in space, which allows us to perform time measurements. Obviously, an unavoidable deadlock appears  if one synchronizes the clocks by assuming a-priori that the one-way speed of light is $c$. In fact, in that case, the one-way speed of light measured with these clocks (that is the Einstein speed of light) cannot be anything else but $c$: this is because the clocks have been set assuming that particular one-way speed in advance. 

Therefore, it can be said that the value of the one-way speed of light is just a matter of convention without physical meaning. In contrast to this, the two-way speed of light, directly measurable along a round-trip, has physical meaning, because round-trip experiments rely upon the observation of simultaneity or non-simultaneity of events at a single point in space and not depends on clock synchronization convention. All well known  methods to measure the speed of light are, indeed, round-trip measurements. The cardinal example is given by the Michelson-Morley experiment: this experiment uses, indeed, an interferometer where light beams are compared in a two-way fashion.

It is important  to emphasize that, consistently with the conventionality of simultaneity, also the value of the speed of light is a matter of convention and has no definite objective meaning.
The constancy of the light velocity in all inertial systems of reference is not a fundamental statement of the theory of relativity. 
The central principle of special relativity is the Lorentz covariance of all  the fundamental laws of physics. Only in Lorentz coordinates the speed of light is independent of the source or observer.

\subsection{Convention-dependent and convention-invariant parts of the dynamics theory}

Consider the motion of charged particle in a given magnetic field.
The  theory of relativity says
that the particle trajectory $\vec{x}(t)$ in the lab frame  depends on the choice of a convention, namely the synchronization convention of clocks in the lab frame. 
Whenever we have a theory containing an arbitrary convention, we should examine what parts of the theory depend on the choice of that convention and what parts do not. We may call the former convention-dependent, and the latter convention-invariant parts. Clearly, physically meaningful  measurement results  must be convention-invariant. 

Consider the motion of two charged particles in a given magnetic field, which is used to control the particle trajectories. Suppose there are two apertures at point $A$ and at point $A'$. From the solution of the dynamics equation of motion we may conclude  that the first particle gets through  the aperture  at $A$ and the second particle gets through the aperture at $A'$ simultaneity. The two events, i.e. the passage of particles at point $A$ and point $A'$  have exact objective meaning i.e. convention-invariant. However, the simultaneity of these two events is convention-dependent and has no exact objective meaning. 
It is important at this point to emphasize that, consistently with the conventionality of simultaneity, also the value of the speed of particle is a matter of convention and has no definite objective meaning.

Many people who learn theory of relativity in the usual way find this disturbing. 
Consider, for instance, the generally accepted experimental fact,
that the magnetic field is only capable of altering the direction of motion, but not the speed of a particle. Since a particle has definite momentum $|\vec{p}|$
there should  be decomposition $|\vec{p}| = mv/\sqrt{1 - v^2/c^2}$ and
the particle  goes along  an arc of a circle with a constant velocity $v$. 
However, we note that this constancy of the speed of the particle is related to the choice of synchronization convention, and actually cannot be subject to experimental tests.

The covariant  particle trajectory is viewed from the Lorentz lab frame as a result of successive infinitesimal Lorentz transformations. 
As one of the consequences of non-commutativity of non-collinear Lorentz boosts, we find an  unusual momentum-velocity relation, which has no  non-covariant analogue. 
The theory of relativity shows us that unusual momentum-velocity relation and Wigner rotation have to do with the effects of acceleration in curved trajectories. 
We point out  that both these effects can be regarded as the two sides of the same coin: they are manifestations of the relativity of simultaneity that is expressed as a  mixture of positions and time. One of the consequences of non-commutativity  of non-collinear Lorentz boosts is a difference between  covariant and non-covariant  single particle trajectories in a constant magnetic field.

In order to examine what parts of the dynamics theory depend on the choice of that convention and what parts do not, we want to show the difference between the notions of path and trajectory. 
So far we have considered the motion of a particle in three-dimensional space using the vector-valued function $\vec{x}(t)$. We have a prescribed curve (path) along which the particle moves. The motion along the path is described by $l(t)$, where $l$ is a certain parameter (in our case of interest the length of the arc).  
The trajectory of a particle conveys more information about its motion because every position is described additionally by the corresponding time instant. The path is rather a purely geometrical notion.  If we take the origin of the (Cartesian) coordinate system and we connect the point to the point laying on the path and describing the motion of the particle, then the creating vector will be a position vector $\vec{x}(l)$.

The difference between path and trajectory is very interesting.
The path $\vec{x}(l)$ has exact objective meaning  i.e. it is convention-invariant. 
In contrast to this, and consistently with the conventionality intrinsic in the velocity, the trajectory $\vec{x}(t)$ of the particle is convention dependent and has no exact objective meaning. 
In order to avoid being to abstract for to long we have given some examples: just think of the experiments related with accelerator physics. Suppose we want to perform a particle momentum measurement. A uniform magnetic field can be used in making a "momentum analyzer" for high-energy charge particles, and it must be recognized that this method for determining the particle's momentum is convention-independent. In fact, the curvature radius of the path in the magnetic field (and consequently the three-momentum) has obviously an objective meaning, i.e. is convention-invariant. Dynamics theory contains a particle trajectory that we do not need to check directly, but which is used in the analysis of electrodynamics problem.

\subsection{Myth about the reality of the time dilation and the length contraction effects}

Generally, experts on the theory of relativity erroneously identify the properties of Minkovski space-time with the familiar form that certain convention-dependent quantities assume under the standard Lorentz coordinatization. These quantities usually are called  "relativistic kinematics effects". There is a widespread belief that the  convention-dependent quantities like the time dilation, length contraction, and Einstein's addition of velocities have direct physical meaning. 
We found that statement like " moving clocks run slow"   is not true under the adopted absolute time clock synchronization, and, hence, are by no means intrinsic features of Minkowski space-time.  
Relativistic kinematic effects are coordinate (i.e. convention-dependent) effects and have no exact objective meaning. In the case of  Lorentz coordinatization, one will experience e.g. the time dilation  phenomenon. In contrast to this, in the case of  absolute time coordinatization there are no relativistic kinematics effects and no time dilation will be found. However, all coordinate-independent quantities like the particle path $\vec{x}(l)$ and momentum  $|\vec{p}|$ remain  independent 
of such a change in clock synchronization.
We demonstrated  that the absolute time synchronization is not artificial: the  accelerator physicists  constantly use it as hidden assumption in their conventional particle tracking codes.  The  difference of the form of  usual relativistic effects in Lorentz and in absolute time coordinatizations will be an important discovery for every  special relativity expert.

\subsection{Reason to prefer the covariant approach}

According to usual accelerator engineering, the study of relativistic particle motion in a constant magnetic field is intimately connected with the old (Newtonian) kinematics: the Galilean vectorial law of addition of velocities is actually used. However, Maxwell's equations are not covariant under   Galilean transformations. We cannot take one kinematics for one part of physical phenomena and the other kinematics for the other, namely Galilean transformations for mechanics and Lorentz transformations for electrodynamics. We must decide which part must be retained and which must be modified.

We demonstrated  in \cite{OURS5,OURS6} that there is no principle difficulty with the non-covariant approach in mechanics and electrodynamics. It is perfectly satisfactory. It does not matter which transformation is used to describe the same reality. What matter is that, once fixed, such convention should be applied and kept in a consistent way in both dynamics and electrodynamics. 
Nevertheless, there is a reason to prefer the covariant approach within the framework of both mechanics and electrodynamics. It is easily seen that the choice of the non-covariant approach also implies the use of much more complicated (anisotropic) electromagnetic field equations.

The common mistake, discussed above, made in accelerator physics is connected with the incorrect algorithm for solving the electromagnetic field equations.   If one wants to use the usual Maxwell's equations, only the solution of the dynamics equations in covariant form (i.e. in Lorentz coordinates)  gives the correct coupling between the Maxwell's equations and particle trajectories in the lab frame.

\section{Space-time and its coordinatization}

The purpose of this section is to present new material concerning the operational foundation of special relativity. To our knowledge, neither operational interpretation of the absolute time coordinatization nor the difference between absolute time synchronization and Einstein's time synchronization from the operational point of view, are given elsewhere in the literature.

\subsection{Coordinatization and operationalism}

Let us discuss an "operational interpretation" of the Lorentz  and absolute time coordinatizations. 
There is a widespread view that only philosophers of physics discuss the issue of distant clock synchronization. Indeed, a  typical physical laboratory contains no space-time grid. 
It should be clear that a rule-clock structure exist only in our mind and manipulations with non existing clocks in the special relativity are an indispensable prerequisite for the application of dynamics and electrodynamics theory in the coordinate representation.
Such situation usually forces  physicist to believe that the application of the  theory of relativity  to the study of physical processes is possible without detailed knowledge of the clocks synchronization procedure.

We should underline that we claim the non covariant approach to relativistic particle dynamics is actually based on the use of a not standard and unusual  clock synchronization assumption within the theory of relativity. The trajectory of the particle, that follows from the solution of the corrected Newton's second law by integrating from initial conditions does not include relativistic kinematics effects. With this radically new factor in the theory of particle dynamics, it is important to know how to operationally interpret the absolute time convention i.e. how  one should perform the clock synchronization  in the lab frame. The result is very interesting, since it tell us about difference between absolute time synchronization and Einstein's time synchronization from the operational point of view.

\subsection{Inertial frame where a source of light is at rest}

Let us give an "operational interpretation" of the Lorentz  coordinatizations. The fundamental laws of electrodynamics are expressed by Maxwell's equations, according to which, as well-known, light propagates with the same velocity $c$ in all directions. This is because Maxwell's theory has no intrinsic anisotropy. It has been stated that in their original form Maxwell's equations are only valid in inertial frames. However, Maxwell's equations can be written down in coordinate representation only if the space-time coordinate system has already been specified.

The problem of assigning Lorentz coordinates  to the lab frame in the case of accelerated motion is complicated. We would like to start with the simpler question of how to assign space-time coordinates to an inertial frame, where a source of light is at rest.  
We need to give a "practical", "operational" answer to this question. The most natural method of synchronization consists in putting all the ideal clocks together at the same point in space, where they can be synchronized. Then, they can be transported slowly to their original places (slow clock transport) \cite{CLOC}.

The usual Maxwell's equations are valid in any inertial frame where sources are at rest and the procedure of slow clock transport is used to assign values to the time coordinate. 
The same considerations apply when charged particles are moving in non-relativistic manner. In particular, when oscillating, charged particles emit radiation, and in the non-relativistic case, when charges oscillate with velocities much smaller than $c$, dipole radiation is generated and described with the help of the Maxwell's equations in their usual form. 

Let's examine in a more detail how the dipole radiation term comes about.
The retardation time in the integrands of the expression for the radiation field amplitude, can be neglected in the cases where the trajectory of the charge changes little during this time. It is easy to find the conditions for satisfying this requirement.  Let us denote by $a$ the order of magnitude of the dimensions of the system. Then the retardation time  $ \sim a/c$. In order to ensure that the distribution of the charges in the system does not undergo a significant change during this time, it is necessary that $a \ll \lambda$, where $\lambda$ is the radiation wavelength. Thus, the dimensions of the system must be small compared to radiation wavelength. This condition can be written  in still another form $v \ll c$, where $v$ is of the order of magnitude of the velocities of the charges. In accounting only for the dipole part of the radiation we neglect all information about the electron trajectory. Therefore, one should not be surprised to find that dipole radiation theory gives fields very much like the instantaneous theory.

The theory of relativity offers an alternative procedure of clocks synchronization based on the constancy of the speed of light in all inertial frames. This is usually considered a postulate but, as we have seen, it is just a convention. The synchronization procedure that follows is the usual Einstein synchronization procedure.  Suppose we have a dipole radiation source. When the dipole light source is at rest, the field equations are constituted by the usual Maxwell's equations. Indeed, in dipole radiation theory we consider the small expansion parameter  $v/c \ll 1$ neglecting terms of order  $v/c$. In other words, in dipole radiation theory we use zero order non relativistic approximation.
Einstein synchronization is defined in terms of light signals emitted by the dipole source at rest, assuming that light propagate with the same velocity $c$ in all directions.  Using  Einstein synchronization procedure in the rest frame of  the dipole source, we actually select the Lorentz coordinate system.

Slow transport synchronization is equivalent to Einstein synchronization in inertial system where the dipole light source is at rest. In other words, suppose we have two sets of synchronized clocks spaced along the $x$ axis. Suppose that one set of clocks is synchronized by using the slow clock transport procedure and the other by light signals. If we would ride together with any clock in either set, we could see that it has the same time as the adjacent clocks, with which its reading is compared. This is because in our case of interest, when light source is at rest, field equations are the usual Maxwell's equations and Einstein synchronization is defined in terms of light signals emitted by a source at rest assuming that light propagates with the same velocity $c$ in all directions.  Using any of these synchronization procedures in the rest frame we actually select a Lorentz coordinate system. In this coordinate system the metric has Minkowski form $ds^2 =  c^2 dt'^2 - dx'^2 - dy'^2 - dz'^2$.
In the rest frame, fields are expressed as a function of the independent variables $x', y', z'$, and $t'$. Let us consider Maxwell's equations in free space.  The electric field $\vec{E}'$ of an electromagnetic wave satisfies the equation $\Box'^2\vec{E}' =  \nabla'^2\vec{E}' - \partial^2\vec{E}'/\partial(ct')^2  = 0$.

\subsection{Motion of a light source with respect to the lab frame}

We  now consider the case when the light source in the lab frame is accelerated from rest up to velocity $v$ along the $x$-axis. A fundamental question to ask is whether our lab clock synchronization method depends on the state of motion of the light source or not. The answer simply fixes a convention. The simplest method of synchronization consists in keeping, without changes, the same set of uniformly synchronized clocks used in the case when the light source was at rest, i.e. we still enforce the clock transport synchronization ( or
Einstein synchronization which is defined in terms of light signals emitted by the dipole source at rest). This choice is usually the most convenient one from the  viewpoint of connection to laboratory reality. 

This synchronization convention preserves simultaneity and is actually based on the absolute time (or absolute simultaneity) convention. After the boost along the $x$ axis, the Cartesian coordinates of the emitter transform as $x' = x-vt, ~ y' = y, ~ z' = z$. This transformation completes with the invariance of simultaneity, $\Delta t' = \Delta t$. The absolute character of the temporal coincidence of two events is a consequence of the absolute concept of time, enforced by $t' = t$. As a result of the boost, the transformation of time and spatial coordinates of any event has the form of a Galilean transformation.  

In the comoving frame, fields are expressed as a function of the independent variables $x', y', z'$, and $t'$. According to the principle of relativity,  the  Maxwell's equations always valid in the Lorentz comoving frame. 
The electric field $\vec{E}'$ of an electromagnetic wave satisfies the equation $\Box'^2\vec{E}' =  \nabla'^2\vec{E}' - \partial^2\vec{E}'/\partial(ct')^2  = 0$.
However, the variables $x',y',z',t'$ can be expressed in terms of the independent variables $x, y, z, t$ by means of a Galilean transformation, so that fields can be written in terms of  $x, y, z, t$. From the Galilean transformation $x' = x - vt, ~ y' = y, ~ z' = z, ~ t' = t $, after partial differentiation, one obtains $\partial/{\partial t} = \partial/{\partial t'} - v\partial/{\partial x'}$, $\partial/{\partial x} = \partial/{\partial x'}$.
Hence the wave equation transforms into

\begin{eqnarray}
&& \Box^2\vec{E} = \left(1-\frac{v^2}{c^2}\right)\frac{\partial^2\vec{E}}{\partial x^2}  - 2\left(\frac{v}{c}\right)\frac{\partial^2\vec{E}}{\partial t\partial x}
+ \frac{\partial^2\vec{E}}{\partial y^2} + \frac{\partial^2\vec{E}}{\partial z^2}
- \frac{1}{c^2}\frac{\partial^2\vec{E}}{\partial t^2} = 0 ~ , \label{GGT2}
\end{eqnarray}

where coordinates and time are transformed according to a Galilean transformation. The solution of this equation $F[x - (c +v)t] + G[x + (-c +v)t]$ is the sum of two arbitrary functions, one of argument $x - (c +v)t$ and the other of argument $x + (-c +v)t$.
Here we obtained the solution for waves which move in the $x$ direction by supposing that the field does not depend on $y$ and $z$. The first term represents a wave traveling forward in the positive $x$ direction, and the second term a wave traveling backwards in the negative $x$ direction. 

We conclude that the speed of light emitted by a  moving source measured in the lab frame $(t, x)$ depends on the relative velocity of source and observer, in our example $v$. In other words, the speed of light is compatible with the Galilean law of addition of velocities. 
In fact, the coordinate velocity of light parallel to the $x$-axis is given by $dx/dt = c + v$ in the positive direction, and $dx/dt = -c + v$  in the negative direction.
The reason why it is different from the electrodynamics constant $c$
is due to the fact that the clocks are synchronized following the absolute time convention, which is fixed because $(t, x)$ is related to $(t',x')$ via a Galilean transformation. 

After properly transforming the d'Alembertian through a Galileo boost, which changes the initial coordinates $(x',y',z',t')$ into $(x,y,z,t)$, we can see that the homogeneous wave equation for the field in the lab frame  has nearly but not quite  the usual, standard form that takes when there is no uniform translation in the transverse direction with velocity $v$. The main difference consists in the crossed term $\partial^2/\partial t\partial x$, which complicates the solution of the equation. To get around this difficulty, we observe that simplification is always possible. The trick needed here is to further make a change of the time variable according to the transformation $t' = t - x v_x/c^2$. In the new variables in i.e. after the Galilean coordinate transformation and the time shift we obtain the  d'Alembertian in the following form

\begin{eqnarray}
\Box^2 =	\left(1-\frac{v_x^2}{c^2}\right)\frac{\partial^2}{\partial x^2} + \frac{\partial^2}{\partial y^2} + \frac{\partial^2}{\partial z^2}
- \left(1-\frac{v_x^2}{c^2}\right)\frac{1}{c^2}\frac{\partial^2}{\partial t^2} ~.
\end{eqnarray}
A further change of a factor $\gamma$  in the scale of time and of the coordinate along the direction of uniform motion leads to the usual wave equation. 

We have, then, a general method for finding solution of electrodynamics problem in the case of the absolute time coordinatization.
Since the Galilean transformation $x = x' + vt', ~ t = t'$, completed by the introduction of the new variables $ct_n = \left[ \sqrt{1-v^2/c^2}ct +  (v/c)x/\sqrt{1-v^2/c^2}\right]$, and
$x_n = x/\sqrt{1 - v^2/c^2}$, is mathematically equivalent to a Lorentz transformation
$x_n = \gamma(x' + vt')~ , t_n = \gamma(t' + vx'/c^2)$, it obviously follows that transforming to new variables $x_n, t_n$ leads to the usual Maxwell's equations.
In particular, when coordinates and time are transformed according to a Galilean transformation followed by the  variable changes specified above, the  d'Alembertian  $\Box'^2 = \nabla'^2 - \partial^2/\partial(ct')^2$  transforms into  $\Box_n^2 = \nabla_n^2 - \partial^2/\partial(ct_n)^2$ . As expected, in the new variables the velocity of light is constant in all directions, and equal to the electrodynamics constant $c$.

The overall combination of Galileo transformation and  variable changes actually yields the Lorentz transformation in the case of absolute time coordinatization in the lab frame, but in this context this transformation are only to be understood as useful mathematical device, which allow one to solve the electrodynamics problem  in the choice of absolute time synchronization  with minimal effort. 

We can now rise an interesting question: do we need to
transform the results of the electrodynamics problem solution  into  the original variables?
We state that the variable changes performed above have no intrinsic meaning - their meaning only being assigned by a convention. 
In particular, one can see the connection between the time shift $t = t' + x'v/c^2$ and the issue of clock synchrony. Note that the final change in the scale of time and spatial coordinates is  unrecognizable also from a physical viewpoint. It is clear that  the convention-independent results of calculations  are precisely the same in the new variables. As a consequence, we should not care to transform the results of the electrodynamics problem solution  into  the original variables.

An idea of studying dynamics and electrodynamics in the case of absolute time coordinatization using technique involving a change of variables is useful from a pedagogical point of view. For example, it is worth remarking that the absent of a dynamical explanation for modulation wavefront rotation  has disturbed some physicists. It should be clear from the preceding discussion that a good way to think of the modulation wavefront rotation is to regard it as a result of transformation to a new time variable.

The question now arises how to operationally interpret these variable changes i.e. how one should change the rule-clock structure of the the lab reference frame. In order to assign a Lorentz coordinate system in the lab frame after the Galilean boost $x = x' + vt', ~ t = t'$, one needs to perform additionally a change scale of reference rules $ x \rightarrow \gamma x$, accounting for length contraction. After this, one needs to change the rhythm of all clocks $t \rightarrow t/\gamma $, thus accounting for time dilation. The transformation of the rule-clock structure completes with the distant clock  resynchronization $t \rightarrow t + xv/c^2$.
This new space-time coordinates in the lab frame are interpreted, mathematically, by saying that the  d'Alembertian  is now diagonal and the speed of light from the moving source  is  isotropic and equal to $c$.

So, from  an operational point of view, the new  coordinates in the lab frame after the clocks resynchronization are impeccable. However, from the theory of relativity we know that if we wish to assign Lorentz coordinates to an inertial lab frame, the synchronization must be defined in terms of light signals. The following important detail of such synchronization  can hardly be emphasized enough.  If the source of light is in motion, we see that the procedure
for distant clocks synchronizing must be performed by using a moving light source.
The constant value of $c$ for the speed  of light emitted by the moving source destroys the  simultaneity introduced by light signals emitted by the (dipole) source at rest. The coordinates reflecting the constant speed of light $c$ from a moving source
are Lorentz coordinates for that particular source.

Consider now two light sources say  "1" and "2". Suppose that in the lab frame the velocities of "1" and "2" are $\vec{v}_1$, $\vec{v}_2$ and $\vec{v}_1 \neq \vec{v}_2$. The question now arises how to assign a time coordinate to the lab reference frame. We have a choice between an absolute time coordinate and a Lorentz time coordinate. The most natural choice, from the point of view of connecting to the laboratory reality, is the absolute time synchronization. In this case simultaneity is absolute, and for this we should prepare, for two sources, only one set of synchronized clocks in the lab frame. On the other hand, Maxwell's equations are not form-invariant under Galilean transformations, that is, their form is different on the lab frame. In fact, the use of the absolute time convention, implies the use of much more complicated field equations, and these equations are different for each source.
Now we are in the position to assign Lorentz coordinates.
The only possibility to introduce Lorentz coordinates in this situation consists in introducing individual coordinate systems  (i.e. individual set of clocks) for each source.
It is clear that if operational methods are at hand to fix the coordinates (clock synchronization in the lab frame) for the first source, the same methods can be used to assign values to the coordinates for the second source and these will be two different Lorentz coordinate systems.

\subsection{Using  different coordinatizations to describe the same physical phenomenon}

In the following we will give some practical examples, in  order not to be too abstract.
There is a realistic configuration encountered in practice, which involves the production of dipole radiation. Light, being a special case of electromagnetic waves, is described by the electrodynamics theory. As we have seen in the foregoing section, the electrodynamics theory meets all requirements of the theory of relativity, and therefore must accurately describe the properties of such a typical relativistic object as light. Let us consider the  case when a dipole light source in the lab frame is accelerated from rest up to  velocity  $v$ along the $x$-axis. 
Consider the effect of light aberration, that is a change in the direction of light propagation ascribed to boosted light sources. The appearance of relativistic effects in radiation phenomena does not depends on of a large speed of of the radiation sources. Lorentz transformations always give rise to relativistic kinematics and no matter how small the ratio $v/c$ may be. 

\subsubsection{Lorentz coordinatization}

The explanation of the effect of aberration of light presented in well-known textbooks is actually based on the use of a Lorentz boost (i.e. of relativistic kinematics) to describe how the direction of a ray of light depends on  the velocity of the light source relative to the lab frame. Let us discuss the special case of aberration of a horizontal ray of light. Suppose that a light source, studied in the comoving frame $S'$, radiates a plane wave along the $z$-axis. Now imagine what happens in the lab frame,  where the source  is moving with constant speed $v$ along the $x$-axis. The transformation of observations from the lab frame  with Lorentz coordinates to the comoving Lorentz frame is described by a transverse  Lorentz boost. On the one hand, the wave equation remains invariant with respect to Lorentz transformations. On the other hand, if make a Lorentz boost, we automatically introduce a time transformation $t' = t - xv/c^2$ and the effect of this transformation is just a rotation of the radiation phase front in the lab frame. 
This is because the effect of this time transformation is just a dislocation in the timing of processes, which has the effect of rotating the plane of simultaneity on the angle $v/c$ in the first order approximation. 
In other words, when a uniform translational motion of the source is treated according to Lorentz transformations, the aberration of light effect is described in the language of relativistic kinematics. In fact, the relativity of simultaneity is a  relativistic effect that appears also in the first order in $v/c$.

\subsubsection{Absolute time coordinatization}

It should be noted, however, that there is another satisfactory way of explaining  the effect of aberration of light. The explanation consists in using a Galileo boost to describe the  uniform translational motion of the light source in the lab frame. After the Galilean transformation of the wave equation we obtain Eq.(\ref{GGT2}).  We are now going to make a  mathematical trick for solution of differential equation with a crossed term:
in order to eliminate the crossed term in the transformed wave equation, we make a change of the time variable $t' = t - xv/c^2$. Using this new time variable  we obtain the wave equation in "diagonal" form, i.e. without crossed term. The time shift results in a slope  of the plane of simultaneity. Then, the electromagnetic waves are radiated at the angle $v/c$, yielding the phenomenon of light aberration. In fact, direction of light propagation is convention-independent and it is precisely the same in the new time variable. As a consequence, we should not care to transform the solution of the electrodynamics problem  into  the original variables.
It should be clear that, in principle, the transformed wave equation may be solved directly without  change of variables, for example by numerical methods, and one may directly derive physical (i.e. convention-invariant) effects associated with the "crossed" term.  The two approaches, treated according to Einstein's or absolute time synchronization conventions give the same result. The choice between these two different approaches is a matter of pragmatics.

\section{Lorentz transformations and Wigner rotation}

\subsection{Lorentz transformations}

Lorentz transformations are essential to the further mathematical development of the Wigner rotation theory, so this section details the usual applications together with some physical discussion.

\subsubsection{The commutativity of collinear  Lorentz boosts}

Let us now consider a relativistic particle, accelerating in the lab frame, and let us analyze its evolution within  Lorentz coordinate systems. The  permanent rest frame of the particle is obviously not inertial and any transformation of observations in the lab frame, back to the rest frame, cannot be made by means of Lorentz transformations. To get around that difficulty one introduces an infinite sequence of comoving frames. At each instant, the rest frame is a Lorentz frame  centered on the particle and moving with it. As the particle velocity changes to its new value at an infinitesimally later instant,  a new Lorentz frame centered on the particle and moving with it at the new velocity is used to observing the particle. 

Let us denote the three inertial frames by $K, R(\tau), R(\tau + d \tau)$. 
The lab frame is $K$, $R(\tau)$  is the rest  frame with velocity $\vec{v} = \vec{v}(\tau)$ relative to $K$, and $R(\tau + d\tau)$ is the rest frame at the next instant of proper time $\tau + d\tau$, which moves relative to $R(\tau)$ with infinitesimal velocity $d\vec{v'}$. 
All inertial reference frames are assumed to be Lorentz reference frames. In order to have this, we impose that $R(\tau)$ is connected to $K$ by the Lorentz boost $L(\vec{v})$, with $\vec{v}$, which transforms a given four vector event $X$ in a space-time into $X_R = L(\vec{v})X$. The relation $X_R = L(d\vec{v'})L(\vec{v})X$ presents a step-by-step change from $K$ to $R(\tau)$ and then to $R(\tau + d\tau)$.

There is another composition of reference-frame transformations which describes the same particle evolution in the Minkowski space-time. 
Let $K(\tau)$
be an inertial frame with velocity $d\vec{v}$ relative to the lab frame $K(\tau + d\tau)$. We impose that $K(\tau)$ is connected to $K(\tau + d\tau)$ by the Lorentz boost $L(d\vec{v})$. The  Lorentz rest frame  $R$ is supposed to move relative to the Lorentz frame $K(\tau)$ with velocity $\vec{v}$. The relation $X_R = L(\vec{v})L(d\vec{v})X$ presents a step-by-step change from $K(\tau + d\tau)$ to  $K(\tau)$ and then to the rest frame $R$. 

Let us examine the transformation of the three-velocity in the theory of relativity. For a rectilinear motion along  the $z$ axis it is performed in accordance with the following equation:
$v_z(\tau + d\tau) = (dv_z'+v_z)/(1 + v_zdv_z'/c^2)$. The "summation" of two velocities is not just the algebraic sum of two velocities, but it is "corrected" by $(1 + v_zdv_z'/c^2)$. 
Like it happens with  the composition of Galilean boosts, collinear Lorentz boosts commute: $L(dv_z)L(v_z) = L( v_z)L(dv_z)$. This means that the resultant of successive collinear Lorentz boosts is independent of which transformation applies first.

\subsubsection{The non-commutativity of two Lorentz boosts in non-parallel directions}

In contrast with the case of Lorentz boosts in collinear directions, Lorentz boosts in different directions do not commute. A comparison with the three-dimensional Euclidean space might help here. 
Spatial rotations do not  commute either. However, also for spatial rotations there is a case where the result of  two successive transformations is independent of their order: that is, when we deal  with rotation around the same axis. While the successive application of two Galilean boosts is Galilean boost and the successive application of two rotations is a rotation,  the successive application of two non-collinear Lorentz boosts is not a Lorentz boost. The composition of non-collinear boosts will results to be equivalent to a boost, followed by spatial rotation, the Wigner rotation. 

Let us compare the succession  $K \to R(\tau) \to R(\tau + d\tau)$ with the succession $K(\tau + d\tau) \to K(\tau) \to R$ in the case when the acceleration in the rest frame is perpendicular to the line of flight of the lab frame in the rest frame.
The frame $R(\tau + d\tau)$ is supposed to move  relative to $R(\tau)$  with velocity $d\vec{v'}_x$.
Because of time dilation in the moving frame, the velocity increment in the lab frame $dv_x$ corresponds to a velocity $\gamma dv_x$ and $-\gamma dv_x$ in the  frames $R(\tau)$ and $R(\tau + d\tau)$ respectively. The resulting boost compositions can be represented as $X_R = L(d\vec{v'}_x)L(\vec{v}_z)X = L( \vec{v}_z)L(d\vec{v}_x)X$. In other words, Lorentz boosts in different direction do not commute: $L( \vec{v}_z)L(d\vec{v}_x) \neq L(d\vec{v}_x)L( \vec{v}_z)$.

Now, since we can write the result in terms of succession $L( \vec{v}_z)L(d\vec{v}_x)$ as well as in terms of succession $L(\gamma d\vec{v}_x)L(\vec{v}_z)$, there is a need to clarify a number of questions associated with these compositions of  Lorentz frames. We can easily understand that the operational interpretation of  the succession  $L( \vec{v}_z)L(d\vec{v}_x)$ is particular simple, involving physical operation used in the measurement of the particle's velocity increment $d\vec{v}_x$ in the lab frame. We should be able to understand the operational interpretation of the succession 
$L(\gamma d\vec{v}_x)L(\vec{v}_z)$. We begin by making an important point: the laws of physics in any one reference frame should be able to account for all physical phenomena, including the observations made by moving observers. The lab observer sees the time dilation in the Lorentz frame which moves with respect to the lab frame with velocity $\vec{v}_z$: $dt/\gamma = d\tau$.  What velocity increment $d\vec{v}_R$ is measured by moving observer? As viewed from the lab frame the moving  observer measures the increment $d\vec{v}_R = \gamma d\vec{v}_x$.

\subsection{Wigner rotation}

\subsubsection{How to measure a Wigner rotation?}

Eq.(1) describes the rotation of the axes of a moving reference frame which is observed in the lab frame. But how to measure this orientation? A moving  coordinate system changes its position in time. The question arises whether it is possible to give an experimental interpretation of the rotation of a moving coordinate system. We illustrate the problem of how to represent orientation of the moving coordinate system with a simple example.

Let us suppose that a  particle moves with velocity $v_z$ along the $z$-axis of a Cartesian $(x,y,z)$ system in the Lorentz lab frame $K$. Let $R$ be the Lorentz comoving frame and designate the Cartesian axes of the comoving frame with $(x',y',z')$, parallel to the Cartesian axes of the  lab frame. How to measure this orientation?   
The natural way to do this is to answer the question: when does each point of the $x'$ axis of the comoving frame cross the $x$-axis of the lab reference system? 
If we have adopted a method for timing distant events (i.e. a synchronization convention), we can also specify a method for measuring the orientation of the comoving frame. For example, if all point of $x'$ axis cross the $x$-axis simultaneously at certain position $z$,  then the $x'$ axis is parallel  to the $x$-axis.

Here we point out that the operations for measuring an orientation of a moving object  are not the same as those for measuring an orientation of a object at rest. We see clearly  that  an absolute significance has been attributed to the concept of simultaneity.

\subsubsection{A simple derivation of the Wigner rotation in ultra-relativistic limit}

The situation relating to the use the Wigner rotation theory in accelerator physics is complicated.
The expression obtained by authors of textbooks 
in the Lorentz lab frame is given by $d\vec{\Phi} =  - (\gamma - 1)\vec{v}\times d\vec{v}/v^2$. 
The theory of relativity shows us that this expression and correct result Eq.(\ref{R}) differ both in sign and magnitude. According to the expression for Wigner rotation in the lab frame presented in textbooks,  the  comoving frame rotates in the opposite direction to the  direction of the velocity vector rotation  and $\Phi = (1- \gamma\theta) \to -\infty$ in the limit $\gamma \longrightarrow \infty$. This contradicts both common sense and, for instance, the  well-known fact that the helicity of a spinning particle is invariant in ultrarelativistic limit. This means that the spin of a massless particle propagating through empty space is always either aligned with the direction of its motion or is opposite to it.

The solution of problems in theoretical physics begins with the application of the qualitative methods. By "qualitative methods" we mean the investigation of limiting cases where one can exploit
the smallness of some parameter. A common mistake of beginners (and also experts who obtained the incorrect expression for Wigner rotation) is to desire to understand everything in its completeness. In real life understanding comes gradually.  
We try to consider the problem in the most simplified form possible.
Rather than working out the  expression for the Wigner rotation  in the case of an arbitrary particle velocity, we would like to demonstrate a method that  allow us to find sign and  magnitude of the Wigner rotation  in the ultrarelativistic limit.

We start with the formulation of the initial conditions in the lab frame in terms of orientation of  comoving coordinate system  and particle velocity. Suppose that  $v_z$ is the velocity of the comoving  frame $R(\tau)$ with respect to the lab frame $K(\tau)$ along  common $z$-axis in positive direction. In the lab frame we select a special type of coordinate system, a 
Lorentz coordinate system to be precise. Within a Lorentz frame (i.e. inertial frame with Lorentz coordinates), Einstein's synchronization of distant clocks and Cartesian space coordinates $(x,y,z)$ are enforced.  In order to have this, we impose that $R$ is connected to $K$ by the Lorentz boost $L(\vec{v}_z)$, with $\vec{v}_z$, which transforms a given four vector event $X$ in a space-time into $X_R = L(\vec{v_z})X$. 

Now we consider the acceleration of the particle in the lab frame up to velocity $dv_x$ along the $x$-axis. The question now arises how to assign synchronization in the lab frame after the particle acceleration. We need to give an "operational" answer to the question how to assign Lorentz coordinates to an inertial lab frame in the case when the particle is accelerated along the $x$-axis. Before particle acceleration  we picked  a Lorentz coordinate system. 
Then, after the acceleration, the particle velocity changes of an small value $dv_x$ along the $x$-axis. 
Without changing synchronization in the lab frame after the particle acceleration, we have complicated situation for electrodynamics of moving charges. As a result of the boost, the transformation of time and spatial coordinates has the form of a Galilean transformation.  
For instance, the electrodynamics problem  is now characterized by anisotropy along the $x$ direction. The main difference arising with respect to Maxwell's equations in their usual form consists in the crossed term $\partial^2/\partial t\partial x$ in Eq.(\ref{GGT2}), which complicates the solution of the electrodynamics equations \footnote{This statement is still valid for any Lorentz covariant field theory}.

In order to keep a Lorentz coordinate system in the lab frame after acceleration, one needs to perform a clock resynchronization by introducing an infinitesimal time shift $t \to t + xdv_x/c^2$. This form of the time transformation is justified by the fact that we are dealing  with first order approximation. Therefore, $dv_x/c$ is so small that $dv_x^2/c^2$ can be neglected and one arrives at coordinate transformation
$x \to x + dv_xt$, $t \to t + xdv_x/c^2$. This infinitesimal Lorentz transformation  differs from a Galilean transformation  only by the inclusion of the relativity of simultaneity, which is the only relativistic effect  appearing in the first order in $dv_x/c$. 
The relation $X_R = L(\vec{v_z})L(\vec{dv_x})X$ presents a step-by-step change from the lab reference frame $K(\tau + d\tau)$ to $K(\tau)$ and then to the proper reference frame $R$.

The shift in the time when points  of the moving axis cross the $x$-axis of the lab frame $\Delta t = xdv_x/c^2$  has, in first order approximation, the effect of  rotation the $x$-axis of the moving coordinate system on the angle $d\Phi = v_z\Delta t/x = v_zdv_x/c^2$ in the first order approximation. 
In the ultrarelativistic limits, $v_z \simeq c$, and the proper reference frame rotates exactly as the velocity vector $\vec{v}$. In vector form this is seen to be
$d\vec{\Phi} =  \vec{v}\times d\vec{v}/v^2$ at $\gamma \to  \infty$.

Let us prove, finally, that the Wigner rotation is a symmetric phenomenon. As well-known, time dilation is symmetric and the same is true for Wigner rotation.
The direction of the velocity rotation in the proper frame  is the same as the direction of the velocity rotation in the lab frame.  We see the result directly from the fact that both velocity and velocity increment  of the lab frame in the proper frame are negative. In fact, $\vec{v}_{R(\tau)} = -\vec{v}_z$ and  $\vec{v}_{R(\tau+ d\tau)} = -\vec{v}_z - \gamma d\vec{v}_x$
The relation $X_R = L(\gamma d\vec{v}_x)L(\vec{v}_z)X$ presents a step-by-step change from the lab reference frame $K$ to  $R(\tau)$ and then to the proper reference frame $R(\tau + d\tau)$.  

Before the  boost along the $x$ direction  we picked  a Lorentz coordinate system in the proper frame.
Then, after the boost, the lab frame velocity changes of a small value $- \gamma dv_x$ along the $x$-axis. 
The question now arises how to synchronize clocks in the proper frame after the boost. 
In order to keep a Lorentz coordinate system in the proper frame after the boost along the $x$- direction, one needs to perform a clock resynchronization by introducing an infinitesimal time shift $\tau \to \tau - x\gamma dv_x/c^2$. The shift in time, when points  of the moving lab axis cross the $x$-axis of the proper frame with $\Delta \tau = - x\gamma dv_x/c^2$,  has the effect of  rotation the $x$-axis of the moving lab coordinate system on the angle $d\Phi_R = - v_z\Delta \tau/x = v_z\gamma dv_x/c^2$ in the first order approximation. In the proper frame the velocity rotation angle would be $\gamma dv_x/v_z$.
In the ultrarelativistic limits, $v_z \simeq c$, and the lab  frame rotates exactly as the velocity vector. In vector form this is seen to be
$d\vec{\Phi}_R =  \gamma \vec{v}\times d\vec{v}/v^2$ at $\gamma \to  \infty$.

The expression for the Wigner rotation angle in the proper frame can be presented in the form $d\vec{\Phi}_R = \vec{v}_R\times d\vec{v}_R/v_R^2$ at $\gamma \to  \infty$. Thus, the Wigner rotation angle in the proper frame is expressed in terms of the lab frame velocity and its increment. This is just the  expression $d\vec{\Phi} =  \vec{v}\times d\vec{v}/v^2$ of the 
Wigner rotation angle in the lab frame at $\gamma \to  \infty$ when rotation angle in the lab frame  is expressed in terms of the proper frame velocity and its increment. So the way to state all this is to say that the Wigner rotation is a symmetric  phenomenon as it must be from kinematic consideration. Although our present discussion is ultrarelativistic, we notice that if we study the motion of a particle moving with an arbitrary velocity, what we deduced about symmetry of the Wigner rotation phenomenon for high velocity is still valid.

Here we only wished to show how naturally Lorentz transformations lead to the Wigner rotation phenomenon. 
We have come to the conclusion that what are usually considered advanced parts of the theory of relativity are, in fact, quite simple. Indeed, we demonstrated that in the ultrarelativistic limit  Wigner rotation results directly from the relativity of simultaneity.
The relativistic kinematics that is involved is particularly simple, involving only infinitesimal Lorentz transformations. 
In this subject we have, of course, the difficulty that the relativistic kinematics is quite strange. The only problem is that we must jump the gap of no longer being able to describe the rotation of a moving object without operational interpretation of such rotation.

\subsubsection{Expression for the Wigner rotation in the case of an arbitrary velocity}

Above we made a simplification in our derivation of the Wigner rotation considering only high velocities. The next question is: what is the general expression for the Wigner rotation? With the results we already have it is a relatively simple to derive it. 

Consider the succession of inertial frame systems $K \to R(\tau) \to R(\tau + d\tau)$.
As viewed from the lab frame the observer in the proper frame measures the velocity increment $d\vec{v}_{R(\tau + d\tau)} = - \gamma d\vec{v}_x$.
The corresponding rotation of the velocity direction in the proper frame $R(\tau + d\tau)$  is $\gamma dv_x/v_z$. In the lab frame  the velocity rotation angle would be $dv_x/v_z$. 
The difference of these two velocity rotation angles $\gamma dv_x/v_z - dv_x/v_z$ is the Wigner rotation angle of the lab frame axises in the proper frame $R(\tau + d\tau)$. One way to see this is follows.

In 3D space we find that the proper frame moves with respect to the lab frame along  the  line motion and the lab frame moves with respect to the proper frame along the same line motion. In other words, it follows that the line motion is the same in the proper frame as in the lab frame.  
The angle between the axis of the  observer's coordinate system and the line motion  is a simple 3D space geometric parameter. The lab observer is able to account for the observation of the rotation angle made in the proper frame and the observation of the rotation angle made in the lab frame.  
Using the line motion as a reference line, the lab observer can then calculate the difference between these angles to find the rotation angle of the lab frame in the proper frame.
We have found that the lab observer sees that the observer in the proper frame measures 
the rotation angle of the lab frame axes  with respect to proper frame axes $\gamma dv_x/v_z - dv_x/v_z$. In vector form this is seen to be  
$d\vec{\Phi}_R = (\gamma - 1)\vec{v}\times d\vec{v}/v^2$ \footnote{The authors of some papers believe that the incorrect result for Wigner rotation in the lab frame presented in textbooks 
$d\vec{\Phi} =  - (\gamma - 1)\vec{v}\times d\vec{v}/v^2$ 	is only incorrectly interpreted  with the understanding that it should be reinterpreted as a Wigner rotation of the lab frame in the proper frame. 	We note that such  reinterpreted expression for Wigner rotation in the proper frame $d\vec{\Phi}  =  - (\gamma - 1)\vec{v}\times d\vec{v}/v^2  \to d\vec{\Phi}_R = - (\gamma - 1)\vec{v}\times d\vec{v}/v^2$ is also incorrect}. 
We wish to remark that the expression for the Wigner rotation angle in the proper frame can be presented in the form $d\vec{\Phi}_R = (1-1/\gamma)\vec{v}_R\times d\vec{v}_R/v_R^2$. 
As we already mentioned the Wigner rotation is a symmetric phenomenon. We can now write expression for the Wigner rotation in the lab frame in the same form
$d\vec{\Phi} = (1-1/\gamma)\vec{v}\times d\vec{v}/v^2$ which corresponds to formula Eq.(\ref{R}).

Why our derivation of the expression for the Wigner rotation is so simple? The reason is that we employed a new method that is very useful in this kind of problem. What we did was to analyze  the physical operations used in the measurement of  Wigner rotation, which has never been done before. 
In fact, the operations for performing measurements on a moving object  are not the same as those for measuring an object at rest, and  an absolute significance has been attributed to the concept of simultaneity. For instance, consider the question: how to describe  Wigner rotation measurement in the proper frame? The easiest way to get the answer is to consider the composition of reference-frame transformations $K \to R(\tau) \to R(\tau + d\tau)$. Similarly, to describe a Wigner rotation measurement in the lab frame, we need to use the composition of reference-frame transformations $K(\tau + d\tau) \to K(\tau) \to R$. 
The standard approach to calculating Wigner rotation in the lab frame is to consider only one succession of inertial frame systems $K \to R(\tau) \to R(\tau + d\tau)$. However a Wigner rotation in the lab frame is described in a more convenient and conceptually clear way by using the succession $K(\tau + d\tau) \to K(\tau) \to R$ rather than the standard one. 

The rotation of axes effect can, of course, be described algebraically  in terms of the transformation matrices for four-vector components.
In textbooks on the theory of relativity, the spatial rotation associated with the composition of two Lorentz boosts in non-parallel directions   is often introduced using the algebraic approach. This is one of the reason why authors of textbooks obtained an incorrect expression for the Wigner rotation. They  describe the rotation of a moving object without operational interpretation of such rotation and encounter serious difficulties in the interpretation of the applied calculations and of the results.

\section{Relativistic particle dynamics}

In previous section we considered the kinematics of the theory of relativity, which concerns the study of the four vectors of positions, velocity and acceleration.   
Kinematics studies  trajectories as geometrical objects, independently of their causes. This means that it is not possible to predict the trajectory of a particle evolving under a given dynamical field using just a kinematic treatment. In dynamics we consider the effect of interaction on motion.
Before treating the case of a spinning charged particle, let us consider the simpler one in which 
we study the motion of a charged particle in electromagnetic fields.

\subsection{Manifestly covariant particle tracking}

Dynamics equations can be expressed as tensor equations in Minkowski space-time. When coordinates are chosen, one may work with components, instead of  geometric objects. Relying on the geometric structure of Minkowski space-time, one can define the class of inertial frames and can adopt a Lorentz frame with orthonormal basis vectors for any given inertial frame. 
In any Lorentz coordinate system the law of motion becomes

\begin{eqnarray}
	&& m\frac{d^2 x_{\mu}}{d\tau^2} = e F^{\mu\nu}\frac{dx_{\nu}}{d\tau}~ ,\label{DDE}
\end{eqnarray}
where here the particle's mass and charge are denoted by $m$ and $e$ respectively.
The electromagnetic field is described by a second-rank, antisymmetric tensor with components $F^{\mu\nu}$. The coordinate-independent proper time $\tau$ is a parameter describing the evolution of physical system under the relativistic laws of motion, Eq. (\ref{DDE}).

The covariant equation of motion for a relativistic charged particle under the action of the four-force $K_{\mu} = e F^{\mu\nu}dx_{\nu}/d\tau$ in the Lorentz lab frame, Eq.(\ref{DDE}), is a relativistic "generalization" of the Newton's second law. 
The three-dimensional Newton second law $md\vec{v}/dt = \vec{f}$  can always be used in the instantaneous Lorentz comoving frame. Relativistic "generalization" means that the previous three independent equations expressing Newton second law are be embedded into the four-dimensional Minkowski space-time.

The immediate generalization of  $md\vec{v}/dt = \vec{f}$ to an arbitrary Lorentz frame is  Eq.(\ref{DDE}), as can be checked by reducing to the rest frame.
In Lorentz coordinates there is a kinematics constraint $u^{\mu}u_{\mu} = c^2$ for the four-velocity $u_{\mu} = dx_{\mu}/d\tau$. Because of this constraint, the four-dimensional dynamics law, Eq.(\ref{DDE}),  actually includes only three independent equations of motion.  
Using explicit expression for Lorentz force we find that the four  equations Eq.(\ref{DDE}) automatically imply the constraint   $u^{\mu}u_{\mu} = c^2$ as it must be.
To prove this,
we calculate the scalar product between both sides of the equation of motion and $u_{\mu}$. Using the fact that  $F^{\mu\nu}$ is antisymmetric (i.e. $F^{\mu\nu} = - F^{\nu\mu}$), we find $u_{\mu}d u^{\mu}/d\tau = eF^{\mu\nu}u_{\mu}u_{\nu} = 0$. Thus, for the quantity $Y = (u^2 - c^2)$ we find $dY/d\tau = 0$.

\subsection{Conventional particle tracking. Hidden absolute time coordinatization}

Having written down the motion equation in a 4-vector form,  Eq.(\ref{DDE}), and determined the components of the 4-force, we satisfied the principle of relativity for one thing, and, for another, we obtained the four components of the equation of particle motion. This is covariant relativistic generalization of the three dimensional Newton's equation of motion which is based on particle proper time as the evolution parameter.

We next wish to describe a particle motion in the Lorentz lab frame using the lab time $t$ as evolution parameter. Let us determine the first three spatial components of the 4-force. We consider for this the spatial part of the dynamics equation, Eq.(\ref{DDE}): $\vec{Q} = (dt/d\tau) d(m\gamma\vec{v})/dt = \gamma d(m\gamma\vec{v})/dt$. The prefactor $\gamma$ arises from the change of the evolution variable from the proper time $\tau$, which is natural since $\vec{Q}$ is the space part of a four-vector, to the lab frame time $t$, which is  needed to introduce the usual force three-vector $\vec{f}$: $\vec{Q} = \gamma\vec{f}$. Written explicitly,  the relativistic form of the three-force is

\begin{eqnarray}
&& \frac{d}{dt}\left(\frac{m\vec{v}}{\sqrt{1-v^2/c^2}}\right) = e\left(\vec{E} + \frac{\vec{v}}{c}\times \vec{H}\right)~ .\label{DDDE1}
\end{eqnarray}

The time component is 

\begin{eqnarray}
&& \frac{d}{dt}\left(\frac{m c^2}{\sqrt{1-v^2/c^2}}\right) = e\vec{E}\cdot\vec{v} ~ .\label{DFE1}
\end{eqnarray}

The evolution of the particle is subject to these four equations, but also to the constraint

\begin{eqnarray}
&& \mathcal{E}^2/c^2 - |\vec{p}|^2 = mc^2  ~ .\label{DCFE1}
\end{eqnarray}

According to  the non-covariant (3+1) approach we seek for the initial value solution to these equations.
Using explicit expression for Lorentz force we find that the three  equations Eq.(\ref{DDDE1}) automatically imply the constraint  Eq.(\ref{DCFE1}),
once this is satisfied initially at $t = 0$.
In the (3+1) approach, the four equations of motion "split up" into (3+1) equations and we have no mixture of space and time parts of the dynamics equation Eq.(\ref{DDE}). 
This approach to relativistic particle dynamics relies on the use of three independent equations of motion  Eq.(\ref{DDDE1}) for three independent coordinates and velocities, "independent" meaning that  equation Eq.(\ref{DFE1}) (and constraint Eq.(\ref{DCFE1})) are automatically satisfied.

One could expect that the particle's trajectory in the lab frame, following from the previous reasoning $\vec{x}(t)$, should be identified with $\vec{x}_{cov}(t)$. However, paradoxical result are obtained by doing so. In particular, the trajectory $\vec{x}(t)$  does not include relativistic kinematics effects.  In the  non-covariant (3+1) approach, the solution of the dynamics problem in the lab frame makes no reference to Lorentz transformations. This means that, for instance, within the lab frame the motion of particles in constant magnetic field looks precisely  the same as predicted  by Newtonian kinematics: relativistic effects do not have a place in this description. In conventional particle tracking  a particle trajectory  $\vec{x}(t)$ can be seen from the lab frame as the result of successive Galileo boosts that track the motion of the accelerated (in a constant magnetic field) particle. The usual Galileo rule for addition of velocities is used to determine the Galileo boosts tracking a particular particle, instant after instant, along its motion along the curved trajectory.

The old kinematics is especially surprising, because we are based on the use of the covariant approach. Where does it comes from?
The previous commonly accepted derivation of the equations for the particle motion in the three dimensional space from the covariant equation Eq.(\ref{DDE}) includes one delicate point. 
In Eq.(\ref{DDDE1}) and Eq.(\ref{DFE1}) the restriction  $\vec{p}  = m\vec{v}/\sqrt{1 - v^2/c^2}$ has already been imposed. One might  well wonder why, because in the accepted covariant approach, the solution of  the dynamics problem for the momentum in the lab frame makes no reference to the three-dimensional velocity. In fact, equation  Eq.(\ref{DDE}) tells us that the force is the rate of change of the momentum $\vec{p}$, but does  not tell us how momentum varies with speed. 
The four-velocity cannot be decomposed into  $u = (c\gamma, \vec{v}\gamma)$  when we deal with a particle accelerating along a curved trajectory in the Lorentz lab frame.  

Actually, the decomposition  $u = (c\gamma, \vec{v}\gamma)$ comes from the relation $u_\mu = dx_\mu/d\tau = \gamma dx_\mu/dt = (c\gamma, \vec{v}\gamma)$. In other words, the presentation of the time component as the relation $d\tau = dt/\gamma$ between proper time and coordinate time is based on the hidden assumption that the type of clock synchronization, which provides the time coordinate $t$ in the lab frame, is based on the use of the absolute time convention. In fact, the calculation carried out in the case of constant magnetic field shows that  $t/\gamma = \tau$  and one can see the connection between this dependence  and the absolute simultaneity convention. Here we have a situation where the temporal coincidence of two events has absolute character:  $\Delta \tau = 0$ implies $\Delta t = 0$.

It will take a long time to become accustomed to the apparently absurd idea that the solution of the corrected Newton's equation by integrating from initial conditions is based on the use the absolute time coordinatization. 
Some of the acknowledged experts in accelerator physics struggled with the absolute time coordinatization not because space-time geometric approach is obscure, but simply because one finds it difficult to outgrow established way of looking at special relativity.

\subsection{Convention-invariant particle tracking}

We already know from our discussion in section 3 that
the path $\vec{x}(l)$ has exact objective meaning  i.e. it is convention-invariant. 
The components of the momentum four vector $mu = (\mathcal{E}/c, \vec{p})$ have also exact  objective meaning. In contrast to this, and consistently with the conventionality intrinsic in the velocity, the trajectory $\vec{x}(t)$ of the particle in the lab frame is convention dependent and has no exact objective meaning.

We want now to describe  how to determine the position vector   $\vec{x}(l)_{cov}$ in covariant particle tracking. We consider the motion in a uniform magnetic field with zero electric field.
Using the Eq.(\ref{DDE}) we obtain

\begin{eqnarray}
	&& \frac{d\vec{p}}{d\tau} =  \frac{e}{mc} ~ \vec{p}\times \vec{H}, ~ ~\frac{d\mathcal{E}}{d\tau} = 0 ~ ~ .
\end{eqnarray}

From $d\mathcal{E}/d\tau = 0$ and from the constraint $\mathcal{E}^2/c^2 - |\vec{p}|^2 = mc^2$   we have $dp/d\tau = 0$, where $p = |\vec{p}| = m|d\vec{x}_{cov}|/d\tau$.
The unit vector $\vec{p}/p$ can be described by the  equation $\vec{p}/p =  d\vec{x}_{cov}/|d\vec{x}_{cov}| = d\vec{x}_{cov}/dl$,
where $|d\vec{x}_{cov}| = dl$ is the differential of the path length.
From the foregoing consideration follows that

\begin{eqnarray}
	&& \frac{d^2\vec{x}_{cov}}{dl^2} =  \frac{d\vec{x}_{cov}}{dl}\times \left (\frac{e\vec{H}}{pc}\right)~ .\label{CPP}
\end{eqnarray}

These three equations corresponds exactly to the equations for the components of the position vector that can be found using the non-covariant particle tracking approach, and $\vec{x}(l)_{cov}$ is exactly equal to $\vec{x}(l)$ as it must be. 

It is interesting to discuss what it means that there are two different (covariant and non covariant) approaches that  produce the same path and particle three-momentum.  The point is that both approaches describe correctly the same  physical reality and since
the curvature radius of the path in the magnetic field, and consequently the three-momentum, has obviously an objective meaning (i.e. is convention-invariant), both approaches yield the same physical results.

\subsection{Phenomenology and relativistic extensions}

In order fully to understand the meaning of the embedding of the Newton's dynamics law in the Minkowski space-time, one must keep in mind that, above, we  characterized  Newton's equation in the Lorentz comoving frame as a phenomenological law. The microscopic interpretation of the inertial mass of a particle is not given. In other words, it is generally accepted that Newton's second law is a phenomenological law  and the rest mass is introduced in an ad hoc manner. The system of coordinates in which the equations of Newton's mechanics are valid can be defined as Lorentz rest frame.  The relativistic generalization of the Newton's second law to any Lorentz frame permits us to make correct predictions.

We are in the position to formulate the following general statement: any phenomenological law, which is valid in the Lorentz rest frame, can be embedded  in the four dimensional space-time only by using Lorentz coordinatization (i.e. Einstein synchronization convention). Suppose we do not know why a muon disintegrates, but we know the law of decay in the Lorentz rest frame. This law would then be a phenomenological law. The relativistic generalization of this law to any Lorentz frame allows us to make a prediction on the average distance traveled by the muon. In particular, when a Lorentz transformation of the decay law is tried, one obtains the prediction that after the travel distance $\gamma v\tau_0$, the population in the lab frame  would be reduced to 1/2 of the origin population. We may interpret this result by saying that, in the lab frame, the characteristic lifetime of a particle has increased from $\tau_0$ to $\gamma\tau_0$. In contrast, in the non covariant (3+1) space and time approach there is no time dilation effect, since for Galilean transformations the time scales do not change.  Therefore,   in the (3+1) non covariant approach, there is no kinematics correction factor $\gamma$ to the travel distance  of relativistically moving muons. The two approaches give, in fact, a different result for the travel-distance, which must be, however, convention-invariant.  This glaring conflict between results of covariant and non covariant approaches can be  explained as follows: it is a dynamical line of arguments that explains this paradoxical situation with the relativistic $\gamma$ factor. In fact, there is a machinery behind the muon disintegration. Its origin is explained in the framework of the Lorentz-covariant quantum field theory. In the microscopic approach to muon disintegration,  Einstein and absolute time   synchronization conventions give the same result for such convention-invariant observables like the average travel distance, and it does not matter which transformation (Galilean or Lorentz) is used.

\subsection{The relativistic mass}

In the non covariant (3+1) space and time approach, there is no time dilation nor length contraction, because for Galilean transformations time and spatial coordinates scales do not change. Moreover, it can easily be verified that Newton's second law keeps its form under Galilean transformations. Therefore, in the (3+1) non covariant approach, there is  no kinematics correction factor $\gamma$ to the mass in Newton's second law. However, in contrast to kinematics effects like  time dilation and length contraction, the correction factor $\gamma$ to the mass in the Newton's second law  has direct objective meaning. In fact, if we assign space-time coordinates to the lab frame using the absolute time convention, the equation of motion is still given by Newton's second law  corrected for the relativistic dependence of momentum on velocity even though, as just stated, it has no kinematical origin. Understanding this result of the theory of relativity is similar to understanding previously discussed results: at first we use Lorentz coordinates and later the (3+1) non covariant approach in terms of a microscopic interpretation that must be consistent with the principle of relativity.


It is well-known from  classical electrodynamics that the electromagnetic field of an electron carries a momentum proportional to its velocity for $v \ll c$, while for an arbitrary velocity $v$, the momentum is altered by the relativistic $\gamma$ factor in the case when the absolute time convention is used. Many attempts have been made to explain the electron mass as fully originating from electromagnetic fields. However, these attempts have failed. In fact,  it is impossible to have a stationary non-neutral charge distribution held together by purely electromagnetic forces. In other words, mass and momentum of an electron cannot be completely electromagnetic in origin and in order to grant stability there is a necessity for compensating electromagnetic forces with non electromagnetic fields. From this viewpoint, Newton's second law is an empirical phenomenological law where the relativistic correction factor $\gamma$ to the mass is introduced in an ad hoc manner.

From a microscopic viewpoint, today accepted explanation of how structureless particles like leptons and quarks acquire mass is based on the coupling to the Higgs field, the Higgs boson having been recently experimentally observed at the LHC. This mechanism can be invoked to explain Newton's second law from a microscopic viewpoint even for structureless particles like electrons. However, at larger scales, an interesting and intuitive concept of the origin of physical inertia is illustrated, without recurring to the Higgs field, by results of Quantum Chromodynamics (QCD)  for protons and neutrons, which are not elementary and are composed of quarks and gluon fields. If an initial, unperturbed nuclear configuration is disturbed, the gluon field generates forces that tend to restore this unperturbed configuration. It is the distortion of the nuclear field that gives rise to the force in opposition to the one producing it, in analogy to the electromagnetic case. But in contrast to the electromagnetic model of an electron, the QCD model of a nucleon is stable,  and other compensation fields are not needed. Now, the gluon field mass can be computed  from the total energy (or momentum) stored in the field, and it turns out that the QCD version in which quark masses are taken as zero provides a remarkably good approximation to reality. Since this version of QCD is a theory whose basic building blocks have zero mass, the most of the mass of ordinary matter (more than 90 percent) arises from pure field energy. In other words, the mass of a nucleon can be explained almost entirely from a microscopic viewpoint, which automatically provides a microscopic explanation of Newton's second law of motion. In order to predict, on dynamical grounds, the inertial mass of a relativistically moving nucleon one does not need to have access to the detailed dynamics of strong interactions. 
It is enough to assume Lorentz covariance (i.e. Lorentz form-invariance of field equations) of the complete QCD dynamics involved in nucleon mass calculations. 

The previous discussion, results in a most general statement: it is enough to assume Lorentz covariance of the quantum field theory involved  in micro-particle (elementary or not elementary) mass calculations
\footnote{Lorentz covariance of the quantum field interactions is an unexplained fact, but all explanation must stop somewhere} in order to obtain the same result for the relativistic mass correction from the two synchronization conventions discussed here,
and it does not matter which transformation (Galilean or Lorentz) is used.

\section{Wigner rotation in the Lorentz lab frame. First practical application}

A very interesting example of Wigner rotation in the Lorentz lab frame is given by the evolution of the modulation wavefront   of an electron beam  in an XFEL. 
Let us suppose that a  modulated electron beam moves along the $z$-axis of a Cartesian $(x,y,z)$ system in the lab frame. As an example, suppose that the modulation wavefront is perpendicular to the velocity $v_z$. How to measure this orientation? A moving electron bunch changes its position with time. The natural way to do this is to answer the question: when does each electron cross the $x$-axis of the reference system? 
If we have adopted a method for timing distant events (i.e. a synchronization convention), we can also specify a method for measuring the orientation of the modulation wavefront: if electrons located  at the position with maximum  density cross the $x$-axis simultaneously at certain position $z$,  then the modulation wavefront is perpendicular to $z$-axis. In other words, the modulation wavefront is defined as a plane of simultaneous events (the events being the arrival of particles located at maximum density): in short, a "plane of simultaneity".

Let us formulate the initial conditions in the lab frame in terms of orientation of the modulation wavefront  and beam velocity. Suppose that  $v_z$ is the velocity of the comoving  frame $R(\tau)$ with respect to the lab frame $K(\tau)$ along the  common $z$-axis in positive direction. In the lab frame we select a special type of coordinate system, a 
Lorentz coordinate system to be precise. Within a Lorentz frame (i.e. inertial frame with Lorentz coordinates), Einstein's synchronization of distant clocks and Cartesian space coordinates $(x,y,z)$ are enforced.  In order to have this, we impose that $R$ is connected to $K$ by the Lorentz boost $L(\vec{v}_z)$, with $\vec{v}_z$, which transforms a given four vector event $X$ in  space-time into $X_R = L(\vec{v_z})X$. 

We now consider the acceleration of the beam in the lab frame up to velocity $v_x$ along the $x$-axis. The question arises how to assign synchronization in the lab frame after the beam acceleration. Before acceleration  we picked  a Lorentz coordinate system. 
Then, after the acceleration, the beam velocity changes of an small value $v_x$ along the $x$-axis. 
Without changing synchronization in the lab frame after the particle acceleration we have a complicated situation as concerns  electrodynamics of moving charges. As a result of such boost, the transformation of time and spatial coordinates has the form of a Galilean transformation.  
In order to keep a Lorentz coordinate system in the lab frame after acceleration, one needs to perform a clock resynchronization by introducing the time shift $t \to t + xv_x/c^2$. This form of time transformation is justified by the fact that we are dealing  with a first order approximation. Therefore, $v_x/c$ is so small that $v_x^2/c^2$ can be neglected and one arrives at the coordinate transformation
$x \to x + v_xt$, $t \to t + xv_x/c^2$. The  Lorentz transformation just described differs from a Galilean transformation  just by the inclusion of the relativity of simultaneity, which is only relativistic effect that appearing in the first order in $v_x/c$. 
The relation $X_R = L(\vec{v_z})L(\vec{dv_x})X$ presents a step-by-step change from the lab reference frame $K(\tau + d\tau)$ to $K(\tau)$ and then to the proper reference frame $R$. 
The shift in the time when  electrons located at the position with maximum density  cross the $x$-axis of the lab frame $\Delta t = xv_x/c^2$  has the effect of a rotation the modulation wavefront on the angle $v_z\Delta t/x = v_zv_x/c^2$ in the first order approximation. 
In ultrarelativistic limits, $v_z \simeq c$, and the modulation wavefront rotates exactly as the velocity vector $\vec{v}$.
Our calculations are performed in ultrarelativistic limit. In the case of an arbitrary electron beam velocity, expression for the Wigner wavefront rotation is given by  Eq.(\ref{R}). 

What does this wavefront readjustment  mean in terms of measurements?  
In the absolute time coordinatization the simultaneity of a pair of events has absolute character. The absolute character of the temporal coincidence of two events is a consequence of the  absolute time synchronization convention. According to this old kinematics, the modulation wavefront remains unvaried. 
However, according to the covariant approach we establish a criterion for the simultaneity of events, which is based on the invariance  of the speed of light. It is immediately understood that, as a result of the motion of electrons along  the $x$ axis (i.e. along the plane of simultaneity before the boost) with the velocity $v_x$, the simultaneity of different events is no longer absolute, i.e. independent of the kick angle $\theta = v_x/c$. This reasoning is in analogy with Einstein's train-embankment thought experiment.   

The wavefront orientation has no exact objective meaning, because the relativity of simultaneity takes place. The statement that the wavefront orientation has objective meaning to within a certain accuracy can be visualized by the picture of wavefront in the proper orientation with approximate angle extension (blurring) given by $\Delta \theta \simeq v_z(v_x/c^2)$. This relation specifies the limits within which the non relativistic  theory can be applied. 
In fact, it follows that for a very non relativistic electron beam for which $v_z^2/c^2$ is very small, the angle "blurring" becomes very small too. In this case angle of wavefront tilt $\theta = v_x/v_z$ is practically sharp $\Delta \theta/\theta \simeq v_z^2/c^2 \ll 1$. This is a limiting case of  non-relativistic kinematics. The angle "blurring" is a peculiarity of relativistic beam motion.  In the ultrarelativistic limit when $v_z \simeq c$, the  wavefront tilt has no exact objective meaning at all  since, due to the finiteness of  the speed of light, we cannot specify any experimental method by which this tilt could be ascertained.

There is a question that we shell try to answer: since in the ultrarelativistic electron beam motion the wavefront orientation not exist at all as physical reality, why do we need to account for the wavefront orientation  in the dynamics calculations?

Let us begin with an example of covariant description:
in the case of Einstein's synchronization convention the covariant electron trajectory is viewed from the Lorentz lab frame as a result of successive infinitesimal Lorentz transformations. The theory of relativity shows  that  covariant trajectories must include the relativistic kinematics effects. When the evolution of the electron beam modulation is treated according to covariant particle tracking, one will experience a Wigner rotation of the beam modulation wavefront: this has no objective meaning   but  is used in the analysis of the electrodynamics problem. 

A comparison with a gauge transformation in  Maxwell's electrodynamics might help here.  There is a reason in favor of using potentials: there are a situations where it seems simpler to solve equations for  potentials $\phi$ and $\vec{A}$ and derive from them the  observable ( gauge invariant) fields rather than to solve the Maxwell's equations for the observable fields. Depending on the choice of  gauge transformation of the electrodynamics potentials   one has different equations. 
The final result of calculations (i.e. observable fields) does not change, but  potentials and equations do, depending on the choice of gauge.  

We have already discussed that the orientation of the modulation wavefront is not a real observable effect; now we have to discuss the observable electrodynamics effects which can be obtained using the modulation wavefront orientation.
If we  couple a charged particle system with electromagnetic fields in accordance with the principle of relativity, we find that  coherent undulator radiation from the modulated electron beam is always emitted in the kicked direction, independently of the system of coordinates (the direction of radiation propagation  has obviously an exact objective meaning). It is like a field calculation problem using potentials. 
This is nothing more than an analogy with the statement  from gauge field theory that potentials (orientation of the modulation wavefront) are not a real observables and that the final result of calculations is gauge-invariant (not depending on the synchronization convention).

It is not difficult to see that coherent undulator radiation from the modulated electron beam is  emitted in the kicked direction 
using a Lorentz coordinate system, where Maxwell's equations are valid and the modulation wavefront is always perpendicular to the beam velocity. In Maxwell's electrodynamics, coherent radiation is always emitted in the direction of the normal to the modulation wavefront. Indeed, we may consider the amplitude of the beam radiated as a whole to be the resultant of radiated spherical waves. This is because Maxwell's theory has no intrinsic anisotropy. The electrons lying on the plane of simultaneity gives rise to spherical radiated wavelets, and these combine according to Huygens'principle to form what is effectively a radiated wave.

We can derive the same results for observables  with the help of Galilean transformations. According to this old kinematics, the orientation of the modulation wavefront remains unvaried. However, Maxwell's equations do not remain invariant with respect to Galilean transformations and the choice of the old kinematics implies the use of anisotropic field equations. In particular, the wave equation for radiated spherical wavelets transforms into Eq.(\ref{GGT2}). The main difference consists in the anisotropic crossed term,  which is of order $v_x/c$. The secondary waves (wavelets) are not spherical, but they are all equal as a consequence of homogeneity. As a result, the wavefront remains plane, but the direction of propagation is not perpendicular to the wavefront. In other words, the radiation beam motion and the radiation wavefront (phase front) normal have different directions. Then, the  Huygens'construction shows that the radiated wave propagates in the kicked direction with the wavefront tilt $v_x/c$.

\section{Relativistic spin dynamics}

\subsection{Magnetic dipole  at rest in an electromagnetic field}

Let us consider at first the spin precession for a non relativistic charge particle. 
The proportionality of magnetic moment $\vec{\mu}$ and angular momentum $\vec{s}$ has been confirmed in many "gyromagnetic" experiments on many different systems. The constant of proportionality is one of the parameters charactering a particular system. It is normally specified by giving the gyromagnetic ratio or $g$ factor, defined by $\vec{\mu} =  ge\vec{s}/(2mc)$. This formula says that the magnetic moment is parallel to the angular momentum and can have any magnitude. For an electron $g$ is very nearly 2.

Suppose  that a particle is at rest in an external magnetic field $\vec{H}_R$. The equation of motion for the angular momentum in its rest frame is  $d\vec{s}/d\tau = \vec{\mu}\times \vec{H}_R =
eg\vec{s}\times\vec{H}_R/(2mc) = \vec{\omega}_s\times \vec{s}$. In other words, the spin precesses around the direction of magnetic field with the frequency $\omega_s = -eg\vec{H}_R/(2mc)$.
In the same non relativistic limit the velocity processes around the direction of $\vec{H}_R$ with the  frequency $\omega_p = -(e/mc)\vec{H}_R$: $d\vec{v}/d\tau = (e/mc)\vec{v}\times\vec{H}_R$. Thus, for $g = 2$ spin and velocity precess with the same frequency, so that the angle between them is conserved.

\subsection{Derivation of the covariant (BMT) equation of motion of spin}

Spin dynamics equations can be expressed as tensor equations in Minkowski space-time. We shell limit ourselves to the case of a particle with a magnetic moment $\vec{\mu}$ in a microscopically homogeneous electromagnetic field. 
Evidently the torque affects only the spin and the force affects only the momentum. It follows that the motion of the system as a whole in any frame is determined entirely by its charge, independent of magnetic dipole moment. This part of the motion has been treated in  the previous  section. We need now only consider the spin motion.

In seeking the equation for the spin motion, we shell be guided  by the known dynamics in the rest frame and the known relativistic transformation laws. We emphasize  that spin is defined in a particular frame (the rest frame). Therefore, to form expressions with known transformation behavior, we need to introduce a four-quantity related to the spin. A convenient choice is a four- (pseudo)-vector $S$ defined by the requirement that in the rest frame its space-like components are the spin components, while the time-like component is zero. We shell call $S$ four-spin; when normalized by dividing by its  invariant length, it will be called  polarization four-vector. It is space-like, and therefore in no frame does it space-like part vanish. 

Let the spin of the particle be represented in the rest frame by $\vec{s}$.
The four-vector $S^{\alpha}$ is by definition required to be purely spatial at time $\tau$ in an instantaneous Lorentz rest frame $R(\tau)$ of the particle and to coincide at this time with the spin $\vec{s}(\tau)$ of the particle; that is $S_R^{\alpha}(\tau) = (0,\vec{S}_R(\tau)) = (0,\vec{s}(\tau))$. At a later instant $\tau + \Delta\tau$ in an instantaneous inertial rest frame $R(\tau +\Delta\tau)$, we have similarly
$S_R^{\alpha}(\tau+\Delta\tau) = (0,\vec{S}_R(\tau+\Delta\tau)) = (0,\vec{s}(\tau+\Delta\tau))$.

The BMT equation is manifestly covariant equation of motion for a four-vector spin $S^{\alpha}$ in an electromagnetic field $F^{\alpha \beta}$:

\begin{eqnarray}
&& \frac{dS^{\alpha}}{d\tau} = \frac{ge}{2mc} \left[F^{\alpha\beta}S_{\beta} + \frac{1}{c^2}u^{\alpha}\left(S_{\lambda}F^{\lambda\mu}u_{\mu}\right) \right] - \frac{1}{c^2}u^{\alpha}\left(S_{\lambda}\frac{du^{\lambda}}{d\tau}\right) ~ ,\label{DDS}
\end{eqnarray}

where $u_\mu = dx_\mu/d\tau$ is the four-dimensional particle velocity vector.
With Eq.(\ref{DDE}), one has  

\begin{eqnarray}
	&& \frac{dS^{\alpha}}{d\tau} = \frac{e}{mc} \left[\frac{g}{2}F^{\alpha\beta}S_{\beta} +  \frac{g-2}{2c^2}u^{\alpha}\left(S_{\lambda}F^{\lambda\mu}u_{\mu}\right) \right]  ~ ,\label{DDSA}
\end{eqnarray}

The BMT equation is valid for any given inertial frame, and consistently describes, together with the covariant-force law, the motion of a charged particle with spin and  magnetic moment.  
If $F^{\mu\nu} \neq 0$, even with $g = 0$, we see that $dS^{\mu}/d\tau \neq 0$. Thus, a spinning charged particle will precess in an electromagnetic field even if it has no magnetic moment.  This precession is pure relativistic effect.

The covariant equation of  spin motion for a relativistic particle under the action of the four-force $Q^{\mu} = e F^{\mu\nu}u_\nu$ in the Lorentz lab frame, Eq.(\ref{DDS}), is a relativistic "generalization" of the equation of motion for a particle angular momentum in its rest frame. Relativistic "generalization" means that the three independent equations expressing the Larmor spin precession are be embedded into the four-dimensional Minkowski space-time. 
The idea of embedding is based on the principle of relativity i.e. on the fact that the classical  equatuion of motion for particle angular momentum $d\vec{s}/d\tau =   eg\vec{s}\times\vec{H}_R/(2mc)$  can always be used in any Lorentz frame where the particle, whose motion we want to describe, is at rest. In other words, if an instantaneously comoving Lorentz frame is given at some instant, one can precisely predict the evolution of the particle spin in this frame during an infinitesimal time interval. 

In Lorentz coordinates there is a kinematics constraint $S^{\mu}u_{\mu} = 0$, which is orthogonality condition of four-spin and four-velocity. Because of this constraint, the four-dimensional dynamics law, Eq.(\ref{DDS}),  actually includes only three independent equations of motion.  Using the explicit expression for Lorentz force we find that the four equations Eq.(\ref{DDS}) automatically imply the constraint  $S^{\mu}u_{\mu} = 0$ as it must be. 
To prove this we may point out that one has in every Lorentz frame $S^0 = \vec{S}\cdot\vec{v}$. While $S^0$ vanishes in the rest frame, $dS^0/d\tau$ need not. In fact $d(S^{\mu}u_{\mu})/d\tau = 0$ implies $dS^0/d\tau = \vec{S}\cdot d\vec{v}/d\tau$. 
The immediate generalization of  $d\vec{s}/d\tau =  eg\vec{s}\times\vec{H}_R/(2mc)$  and 
$dS^0/d\tau = \vec{S}\cdot d\vec{v}/d\tau$
to arbitrary Lorentz frames is Eq.(\ref{DDS}) as can be checked by reducing to the rest frame.
A methodological analogy with the relativistic generalization of the Newton's second law emerges.

In order to fully  understand the meaning of  embedding of the spin  dynamics law in the Minkowski space-time, one must keep in mind that, above, we  characterized the spin dynamics equation in the Lorentz comoving frame as a phenomenological law. The microscopic interpretation of the magnetic moment of a particle is not given. In other words, it is generally accepted that the spin dynamics law is a phenomenological law  and the magnetic moment is introduced in an ad hoc manner. The system of coordinates in which the classical equations of motion for particle angular momentum are valid can be defined as Lorentz rest frame.  The relativistic generalization of the three-dimensional equation  $d\vec{s}/d\tau =  eg\vec{s}\times\vec{H}_R/(2mc)$ to any Lorentz frame permits us to make correct predictions.

\subsection{Change spin variables}

The equation  Eq.(\ref{DDSA}) is more complex than one might think. In fact, it is composed by a set of coupled differential equations.  To find solution directly from the system seems quite difficult, even for a very symmetric, uniform magnetic field setup.

In order to apply Eq.(\ref{DDSA}) to specific problems it is convenient to introduce a three vector $\vec{s}~$ by the equation $\vec{s} = \vec{S} + S^{0}\vec{p}c/(\mathcal{E} +  mc^2 )$. 
With the help of this relation one can work out the equation of motion for $\vec{s}~$. In the important case  of a uniform magnetic field with no electric field in the lab frame one has, after a somewhat lengthly calculations: 

\begin{eqnarray}
	&& \frac{d\vec{s}}{d\tau} = - \frac{e}{2m} \left[\left(g-2 + \frac{2}{\gamma}\right)\gamma\vec{H} - \frac{(g-2)\gamma}{\gamma +1}\frac{\vec{v}}{c}\left(\frac{\vec{v}}{c}\cdot\gamma\vec{H}\right)\right]\times\vec{s}  ~ ,\label{DDSS}
\end{eqnarray}

What must be recognized is that in the accepted covariant approach (indeed, Eq.(\ref{DDSA}) is obviously manifestly covariant), the solution of the dynamics problem for the spin in the lab frame makes no reference to the three-dimensional velocity. In fact, the Eq.(\ref{DDSS}) includes  relativistic factor $\gamma$  and vector $\vec{v}/c$, which are actually notations:  $\gamma = \mathcal{E}/(mc^2)$,   $\vec{v}/c = \vec{p}c/\mathcal{E}$. All quantities $\mathcal{E}, \vec{p}, \vec{H} $ are measured in the lab frame and have exact objective meaning i.e. they are convention-independent. The evolution parameter $\tau$ is also  measured in the lab frame and has exact objective meaning .  For instance, it is not hard to demonstrate that $d\tau = m dl/|\vec{p}|$, where $dl$ is the differential of the path length.

Spin vector  $\vec{s}~$ is not part of a four-vector,  and depends on both $\vec{S}$ and $S^{0}$. While not being a four-vector, it is effectively a three-dimensional object (having zero time component  in the inertial frame in question) and  the spatial part of this object undergoes pure rotation with constant rate for the example of motion along a circle in special relativity.
If we perform an arbitrary velocity mapping, $\vec{s}$ will have to be recomputed from the transformed values $S^{\mu}$ and $p^{\mu}$. However, this new $\vec{s}$ will satisfy an equation of the form  Eq.(\ref{DDSS}), with $\vec{H}$  computed from the transformed $F^{\mu\nu}$.

Let us restrict our treatment of spinning particle dynamics to purely transverse magnetic fields. 
This means that the  magnetic field vector $\vec{H}$ is oriented normal to the particle line motion. If the field is transverse, then    equation Eq.(\ref{DDSS}) is reduced to

\begin{eqnarray}
	&& \frac{d\vec{s}}{d\tau} = \vec{\Omega}\times\vec{s} = - \frac{e}{2mc} \left[\left(g-2 + \frac{2}{\gamma}\right)\gamma\vec{H} \right]\times\vec{s}   ~ ,\label{DDSS1}
\end{eqnarray}

Now we have an equation in the most convenient form to be solved. Suppose we let the charged spinning particle in the lab frame through a bending magnet with the length $dl$. We know that $d\theta = - eHdl/(|\vec{p}|c)$ is the orbital angle of the particle in the lab frame. Note that $d\tau = mdl/|\vec{p}|$. Then, Eq.(\ref{DDSS1}) tells us that 
we may write  the spin rotation angle  with respect to the lab frame axes $\Omega d\tau$ as $ \Omega d\tau = [(g/2 - 1)\gamma d\theta  + d\theta]$. 

This tell us  that in the lab frame the spin of a particle  $\vec{s}$ changes the angle $\phi$ with its line motion.  The rate of change of the angle $\phi$ with the orbital angle is  $(g/2 - 1)\gamma$, so we can write $d\phi = (g/2 - 1)\gamma d\theta $. 

We would like to discuss the following question: Is the vector $\vec{s}$ merely a device which is useful in making calculations - or is it a real quantity ( i.e. a quantity which has direct physical meaning)?
Knowing that there is a simple algebraic relation between $\vec{s}$ and the standard spin vector, the  spin vector $\vec{s}$  can be used as an intermediate step to easily find the standard spin vector $S^{\mu}$. There is, however, also a direct physical meaning to the spin vector $\vec{s}$ . The spin vector $\vec{s}$ directly gives the spin as perceived in a comoving system.

The approach in which we deal with the proper spin is much preferred in the experimental practice due to mathematical simplicity and clear physical meaning of the vector $\vec{s}$. 
Unlike momentum, which has definite components in each reference frame, angular momentum is defined only in one particular reference frame. It does not transform. Any statement about it refers to the rest frame as of that instant. If we say that in the lab frame the spin of a particle makes the angle $\phi$ with its velocity, we mean that in the particle's rest frame the spin makes this angle with the line motion  of the lab frame.

\subsection{An alternative approach to the BMT theory}

\subsubsection{BMT equation transformed to the rest frame}

When Bargman, Michel and Telegdi first discovered the correct laws of spin dynamics, they wrote 
a  manifestly covariant  equation in Minkowski space-time, Eq.(\ref{DDS}), which describes the motion of the four spin $S^{\mu}$. The derivation of this equation was  very similar to the four-tensor equations that were already known to relativistic particle dynamics. How to solve this four-tensor equation is an interesting question. In relativistic spin dynamics it is done  in one particular way, which is very convenient.
In order to apply four-tensor equation   Eq.(\ref{DDS})
to specific problems  it is convenient to transform this equation to the rest frame as of that instant. Should one be surprised that the starting point of Bargman, Michel and Telegdi  was the particle rest frame and the classical equation of motion for particle angular momentum, which they generalized  to the Lorentz lab frame and then transformed  back to the rest frame?

We want to emphasize that the equation Eq.(\ref{DDSS1}) for the proper spin  $\vec{s}$ and the BMT equation Eq.(\ref{DDS}) for the four spin  $S^{\mu}$ are completely equivalent, they both determine the behaviour of the spin from the point of view of the lab frame. With  Eq.(\ref{DDSS1}) have what we need to know - the evolution  of the proper spin vector  $\vec{s}$ with respect to the lab frame axes. Starting  from the classical equation  $d\vec{s}/d\tau =  eg\vec{s}\times \vec{H}_R/(2mc)$, which describes the Larmor precession with respect to the proper frame axes, we have derived the equation Eq.(\ref{DDSS}), which describes the spin motion with respect to the lab frame axes in the proper frame and reduced  to  Eq.(\ref{DDSS1}) in the case of purely transverse magnetic fields. That means that we know the orientation of the proper spin with respect to the lab coordinate system which is moving with velocity $-\vec{v}$ and acceleration $-\gamma d\vec{v}/d\tau$ in the proper frame.

Above we described the BMT equation, Eq.(\ref{DDSS1}), in the standard manner. It uses a spin quantity defined in the proper frame but observed with respect to the lab frame axes.
Let's look at what the equation Eq.(\ref{DDSS1}) says in a little more detail. It will be more convenient if we rewrite this equation as

\begin{eqnarray}
	&& d\vec{s} =  \vec{\Omega} d\tau \times\vec{s} =  -eg\gamma \vec{H}d\tau/(2mc)\times\vec{s}
+  e(\gamma - 1) \vec{H}d\tau/(mc)\times\vec{s} ~ .\label{DDSS11}
\end{eqnarray}

\subsubsection{Relativistic kinematic addition to the Larmor rotation}

Now let's see how we can write  the right-hand side of Eq.(\ref{DDSS11}) . 
The first term is that we would expect for the spin rotation due to a torque with respect to the proper frame axes $d\vec{\phi}_L = -eg\gamma \vec{H}d\tau/(2mc) = (g/2)\gamma d\vec{\theta}$.
Here $d\vec{\theta} = -eHdl/(|\vec{p}|c)$ is the angle of the velocity rotation  in the lab frame. 
It has also been made evident by our analysis in the previous Section 4 that angle of rotation $d\vec{\phi}_W =  -e(\gamma - 1) \vec{H}d\tau/(mc) = (\gamma - 1)d\vec{\theta}$ 
corresponds to the Wigner rotation of the lab frame axes with respect to the proper frame axes. 
With this definitions, we have


\begin{eqnarray}
&& d\vec{s} =  \vec{\Omega} d\tau \times\vec{s} = d\vec{\phi}_L \times\vec{s} -  d\vec{\phi}_W \times\vec{s} ~ ,\label{DDSS2}
\end{eqnarray}

which begins to look interesting.

\subsubsection{Wigner rotation in the proper frame. First practical application}

Now we introduce our new approach to the BMT theory, finding 
another way in which our complicated problem can be solved.
We know  that $d\vec{\phi}_L$ and $d\vec{\phi}_W$ are the rotations with respect to the proper frame axes. Actually we only need  to find the spin motion with respect to the lab frame axes. Now we must be careful about signs of rotations.

There is a good mnemonic rule   to learn the signs of  different rotations. The rule says, first, that the direction of the velocity rotation in the proper frame is the same as the direction of the velocity rotation in the lab frame. Second, the direction of the lab frame rotation in the proper frame  is the same as the direction of the velocity rotation in the proper frame.
Third, the sign of the spin rotation due to a torque  at $g > 0$ (it is handy to  remember that for an electron $g$ is positive and very nearly 2) means that the direction of the rotation in the proper frame is the same as the direction of the velocity rotation in the proper frame.

We now ask about the proper spin rotation with respect to  to the lab frame axes. 
This is easy to find. The relative rotation angle is $d\vec{\phi}_L - d\vec{\phi}_W $. 
So we begin to understand the basic machinery behind  spin dynamics. 
We see why the Wigner rotation of the lab frame axes in the proper frame must be taken into account if we need to know the proper spin dynamics with respect to the lab frame axes.

Why the new derivation of the BMT equation is so simple? The reason is that the splitting of the particle spin motion with respect to the lab frame axes into the  dynamic (Larmor) and kinematic (Wigner) parts can only be realized in the proper frame. In the proper frame, we do not need to know any more about a relativistic "generalization" of the (phenomenological) classical equation of motion for the particle angular momentum. 
In this case, it is possible to separate  the spin dynamics  problem into the trivial  dynamic problem and into the kinematic problem of  Wigner rotation of the lab frame in the proper frame.

\subsection{Spin tracking}

Having written down the spin motion equation in a 4-vector form, Eq.(\ref{DDSA}), and determined the components of the 4-force, we satisfied the principle of relativity for one thing, and, for another, we obtained the four components of the equation for the spin motion. This is a covariant relativistic generalization of the usual three dimensional equation of magnetic moment motion, which is based on the particle proper time as the evolution parameter.
We next wish to describe the spin motion with respect to  the Lorentz lab frame using the lab time $t$ as the evolution parameter. 

\subsubsection{Conventional spin tracking. Hidden absolute time coordinatization}

When going from the proper time $\tau$ to the lab time $t$, the frequency of spin precession with respect to the lab frame can be obtained  using the well-known formula $d\tau = dt/\gamma$. We then find

\begin{eqnarray}
&& \frac{d\vec{s}}{dt} = \vec{\varpi}\times\vec{s} = - \frac{e}{2mc} \left[\left(g-2 + \frac{2}{\gamma}\right)\vec{H} \right]\times\vec{s}   ~ .\label{DDSS3}
\end{eqnarray}

The frequency of spin precession can be written in the form 

\begin{eqnarray}
&& \varpi = \omega_0[1 + \gamma (g/2 - 1)]  ~ ,\label{DDSS5}
\end{eqnarray}

where $\omega_0$ is the particle revolution frequency.
Now  the time-like part of the four-velocity  is decomposed to $c\gamma = c/\sqrt{1 - v^2/c^2}$ and the trajectory does not include relativistic kinematics effects. In particular, the Galilean vectorial law of addition of velocities is actually used. So we must have made a jump to the absolute time coordinatization.

The previous  commonly accepted derivation of the equations for the spin precession in the lab frame  from the covariant equation Eq.(\ref{DDS}) has the same delicate point as the  derivation of the equation of particle motion from the covariant equation  Eq.(\ref{DDE}). 
The four-velocity cannot be decomposed into  $u = (c\gamma, \vec{v}\gamma)$  when we deal with a particle accelerating along a curved trajectory in the Lorentz lab frame.  One of the consequences 
of non-commutativity of non-collinear Lorentz boosts is the unusual momentum-velocity relation. In this case there is a difference between covariant and non-covariant particle trajectories. 

The old kinematics comes from the relation $d\tau = dt/\gamma$. 
The presentation of the time component simply as  the relation $d\tau = dt/\gamma$ between proper time and coordinate time is based on the hidden assumption that the type of clock synchronization that  provides the time coordinate $t$ in the lab frame is based on the use of the absolute time convention.

\subsubsection {Convention-invariant spin tracking}

In the last Section we saw that the particle path $\vec{x}(l)$ has an exact objective meaning i.e. it is convention-invariant. The spin orientation $\vec{s}$  at each point of the  particle path $\vec{x}(l)$  has also exact objective meaning. In contrast to this, and consistently with the conventionality of the three-velocity, the function $\vec{s}(t)$ describing the spinning particle  in the lab frame has no exact objective meaning. 

We now want  to describe how to determine the spin orientation  along the path $\vec{s}(l)$ in covariant spin tracking. Using the covariant equation Eq.(\ref{DDS}) we obtain Eq.(\ref{DDSS1}). If we use the  relation $d\tau = mdl/|\vec{p}|$ our  convention-invariant equation of spin motion  reads

\begin{eqnarray}
&& \frac{d\vec{s}}{dl}  = - \frac{e\mathcal{E}}{m|\vec{p}|c^3} \left[\left(\frac{g}{2}-1 + \frac{mc^2}{\mathcal{E}}\right)\vec{H} \right]\times\vec{s} 
= \left[\left(\frac{g}{2} - 1\right)\frac{\mathcal{E}}{mc^2} + 1\right]\frac{d\vec{\theta}}{dl}\times\vec{s} ~ ,\label{DDSS4}
\end{eqnarray}

which is based on the path length $l$ as the evolution parameter.  
These three equations corresponds exactly to the equations for components of the proper spin vector  that can be found from the non-covariant spin tracking equation  Eq.(\ref{DDSS3}).  
So everything comes out all right.
We want to emphasize that there are two different (covariant and non covariant) approaches that  produce the same spin orientation $\vec{s}(l)$ along the path.  The point is that both approaches describe correctly the same  physical reality and the orientation of the proper spin $\vec{s}$ at any point of particle path in the magnetic field has obviously an objective meaning, i.e. is convention-invariant. 

Now we take an example, so it can be seen that 
we do not need to ask questions about the function $\vec{s}(t)$ of a spinning particle experimentally. Just think of  experiments related with accelerator physics. 
Suppose we want to calibrate  the  beam energy in a storage ring based on measurement of spin precession frequency of polarized electrons. To measure the precession frequency $\varpi$, a method of beam resonance  depolarization by an oscillating  electromagnetic field can be used. 
There are many forms of depolarizers, but we will mention just one, which especially simple. It is a depolarizer whose operation depends on the radio-frequency longitudinal magnetic field which is produced by a current-curring loop around a ceramic section of the vacuum chamber. 

Suppose the observer in the lab frame performs the  beam energy measurement. We should examine what parts of the measured data depends on the choice of synchronization convention and what parts do not. Clearly, physically meaningful results must be convention-invariant.
One might think that this is a typical time-depending measurement of function $\vec{s}(t)$. However, we state that the precession frequency  $\varpi$ has no intrinsic meaning - its meaning is only being assigned by a convention. It is not possible to determine the precession frequency $\varpi$  uniquely, because there is always  some arbitrariness in the $\vec{s}(t)$. For instance, it is always possible to make an arbitrary change in the rhythm of the clocks (i.e. scale of the time).  But our problem is to determine the energy for an electron beam. So one needs to measure also the revolution frequency $\omega_0$ by using the same space-time grid. What this all means physically is very interesting. The ratio
$\varpi/\omega_0$ is convention independent i.e it does not depend on the distant clocks synchronization or on the rhythm of the clocks.
It means, for example, that if we  observe  the dimensionless frequency $ \varpi/\omega_0$, 
we can find out the value of the convention-invariant beam energy $\mathcal{E}$. The $(g/2 -1)$ factor can be calculated by use of quantum electrodynamics.

Let us now return to our examination  of the measured data in experiments related with the calibration of the beam energy in a storage ring. The spin $\vec{s}$ of a particle makes the angle $\phi$  with it velocity. From Eq.(\ref{DDSS4})  we have been able to write 
the angle $\phi$ in therm of orbital angle $\theta(l)$  in a form  $\phi = \phi(\theta)$. We thus use the orbital angle $\theta$ as evolution parameter.  Suppose that the depolarizer is placed at an azimuth $\theta_0$.
During a period of velocity rotation, the spin will rotate through an angle of $\Delta \phi = \phi(\theta_0 + 2\pi) - \phi(\theta_0)$. The point is that depolarizer measurements are made to determine the observable $\Delta \phi$.
Let us see how  equation Eq.(\ref{DDSS4}) gives the observable $\Delta \phi$.
It can be written in integral form  $\Delta \phi  = \int d\theta [(g/2 -1)\mathcal{E}/(mc^2)] = 2\pi[(g/2 -1)\mathcal{E}/(mc^2)]$. We can already conclude something from these results. The convention-invariant observation $\Delta \phi$ is actually a geometric parameter.   
It comes quite naturally that in experiments related with spin dynamics in a storage ring we do not need to ask question  about the function $\vec{s}(t)$ experimentally.

\subsection{Spin rotation in the limit $g \to 0$  as dynamics effect} 

\subsubsection{Spin tracking at $g \to 0$}

It is generally accepted that spin dynamics law is a phenomenological law  and that the magnetic moment is introduced in an ad hoc manner. 
Let us consider the special case with $g \to 0$.
The BMT equation for a particle with small $g$ factor is

\begin{eqnarray}
	&& \frac{dS^{\alpha}}{d\tau} =  - \frac{1}{c^2}u^{\alpha}\left(S_{\lambda}\frac{du^{\lambda}}{d\tau}\right)
	= - \frac{e}{mc^3}u^{\alpha}\left(S_{\lambda}F^{\lambda\mu}u_{\mu}\right) ~ .\label{DDS6}
\end{eqnarray}

It is often more convenient to write this equation as the equation of motion for $\vec{s}$.
If the field is transverse, then the equation Eq.(\ref{DDS6}) is reduced to
 
\begin{eqnarray}
	&& \frac{d\vec{s}}{d\tau} =  \left[\left(\frac{\mathcal{E}}{mc^2} -1\right)\frac{e}{mc}\vec{H} \right]\times\vec{s}   ~ ,\label{DDSS7}
\end{eqnarray}

Note that the equation Eq.(\ref{DDSS7}) for the proper spin  $\vec{s}$ and the BMT equation Eq.(\ref{DDS6}) for four spin  $S^{\mu}$ are completely equivalent. Eq.(\ref{DDSS7}) is the result of transformation to new spin variables.

Conventional spin tracking treats the space-time continuum in a non relativistic format, as a (3+1) manifold. In the conventional spin tracking, we assign absolute time coordinate and we have no mixture of positions and time. This approach to relativistic spin dynamics relies on the use of three  equations for the spin motion

\begin{eqnarray}
&& \frac{d\vec{s}}{dt} =  \left[\left(1 -  \frac{1}{\gamma}\right) \frac{e}{mc} \vec{H} \right]\times\vec{s} 
= - \left[\left( \gamma - 1\right) \vec{\omega_0} \right]\times\vec{s} ~ ,\label{DDSS8}
\end{eqnarray}

which  are based on the use of the absolute time $t$ as the evolution parameter. Here, $\vec{\omega_0} = -e\vec{H}/(mc\gamma)$ is the particle angular frequency in the lab frame. Now  the time-like part of the four-velocity  is decomposed to $c\gamma = c/\sqrt{1 - v^2/c^2}$. This decomposition is a manifestation of the absolute time convention.

There are two different (covariant and non covariant) approaches that  produce the same spin orientation $\vec{s}(l)$ along the path.  
Using the Eq.(\ref{DDSS7}) or Eq.(\ref{DDSS8}) we obtain

\begin{eqnarray}
&& \frac{d\vec{s}}{dl} =  - \left[\frac{\mathcal{E}}{mc^2} - 1\right]\frac{d\vec{\theta}}{dl}\times\vec{s} ~ ,\label{DDSS9}
\end{eqnarray}

Both approaches describe correctly the same  physical reality, and the rotation of the proper spin $\vec{s}$ with respect to the lab frame axes  at $g \to 0$ is convention-invariant.

\subsubsection{Origin of  spin rotation at $g \to 0$}

It is generally believed that "If the particle with spin has no magnetic moment  ($g = 0$), the spin precession is purely kinematic in nature ...  For example, the $~^{235}_{92}$U nucleus has a rather small $g$ factor ($g = - 0.26$), which is smaller than that of the normal gyro-magnetic ratio by a factor 8. For such object, the kinematic effect dominates over the dynamics one."\cite{ST}. 

This statement  presented in most published papers and textbooks is incorrect and misleading.  
The reason is that a splitting of particle spin motion into the dynamic and kinematic (Wigner) parts cannot be performed in the Lorentz lab frame. In the Lorentz lab frame, Eq.(\ref{DDS6}), is a relativistic "generalization" of the equation of motion for a particle angular momentum in its rest frame. In other words, Eq.(\ref{DDS6}) is a dynamics equation.

The relativistic kinematic effects such as Wigner rotation, Lorentz-Fitzgerald contraction, time dilation and relativistic corrections to the law of composition of velocities are coordinate (i.e. convention-dependent) effects and have no exact objective meaning. In the case of the Lorentz coordinatization, one will experience e.g. the Wigner rotation phenomenon. In contrast to this, in the case of  absolute time coordinatization there are no relativistic kinematics effects, and no Wigner rotation will be found.  

However, the spin orientation at each point of the particle path has exact objective meaning. In fact, Eq.(\ref{DDSS9}) is convention-invariant i.e includes only quantities which have exact objective meaning.
Understanding this result of the theory of relativity is similar to understanding the previously discussed result for relativistic mass correction. We find that  the evolution of a particle along its path  is still given by the corrected Newton's second law  even though the relativistic mass correction has no kinematical origin. 
A methodological analogy with the spin dynamics equation  Eq.(\ref{DDSS9})    emerge by itself. The spin rotation in the lab frame at $g \to 0$ is relativistic effect (as the relativistic mass correction) but it has no kinematical origin.

In relation with this discussion, we would now like to describe an apparent paradox. The spin of a particle with small $g$ factor  experiences no torque in the proper frame. However in the lab frame the spin experiences torque from the magnetic part of the force. How can there be a torque and so a time rate of change of angular momentum  in one inertial frame and no torque in another?  Is there a paradox? In the considered case many physicists  would like to think that the principle of relativity is violated. Nature apparently doesn't see the paradox, however, because
we discuss the different setups in a different inertial frames. In fact,  the  spinning particle is at rest in one inertial frame and is moving in another. Let us now discuss more of the consequences of the principle of relativity.
It says that the laws of nature are the same (or take the same form) in all inertial frames. For instance, in agreement with this principle, usual Maxwell's equations can always be exploited in any inertial frame where electromagnetic sources are at rest using Einstein synchronization procedure in the rest frame of the source. Now the question is, what is the restriction in our case of interest? 
Suppose we try a $~^{235}_{92}$U nucleus, which is at rest in the  lab frame. In agreement with the principle of relativity, 
the  spin of the $~^{235}_{92}$U nucleus  will  experience no torque from the magnetic part of the force in this case. That is the meaning of the principle of relativity in this example: it means that all experiments performed in the moving frame with this nucleus,  and all phenomena in the moving frame will appear the same as if the  frame were not moving.

\subsubsection{Lab frame view of the spin rotation at $g \to 0$}

Suppose  we do not know why the nucleus $~^{235}_{92}$U has its anomalous magnetic moment, but we know the law of spin precession in the Lorentz rest frame. This law would then be a phenomenological law. The relativistic generalization of this law to any Lorentz frame allows us to make a prediction on the spin rotation in the Lorentz lab frame using convention independent equation Eq.(\ref{DDSS9}). 

However, as already discussed, there is another satisfying way to describe the same experimental setup based on the absolute time convention.  It is a dynamical line of arguments that explains the spin rotation in the lab frame. In fact, we know that there is a machinery behind the spin rotation. Its origin is explained in the framework of the Lorentz-covariant  quantum field theory, the well-known Quantum Chromodynamics (QCD). In the microscopic approach to our nucleus spin rotation in the lab frame,  Lorentz and absolute time coordinatizations give the same result for such convention-invariant observables like the spin rotation in the lab frame, and it does not matter which coordinatization is used.   

Let us examine in a little more detail how this spin rotation  comes about from a microscopic viewpoint.
The explanation consists in using the absolute time coordinatization and  a Galileo boosts to describe a spinning particle accelerating along a curved trajectory in the lab frame. After the Galilean transformations of the quantum field equations we would obtain  complicated (anisotropic ) QCD equations.  The new terms that have to be put into the field equations due to the use of Galilean transformations lead to the same prediction as concerns experimental results: the spin of the nucleus is rotated with respect to the lab frame axes according to Eq.(\ref{DDSS9}).

How shell we solve the quantum field equations after the Galilean transformations? It is like an electromagnetic problem with light aberration. It is enough to assume Lorentz covariance of the quantum field theory involved  in the nucleus magnetic moment calculations. 
We are  can make a  mathematical trick for the solution of the quantum field equations with  anisotropic terms: in order to eliminate these terms in the transformed field equations, we make a change of  variables. Using new  variables  we obtain the phenomenon of spin rotation with respect to the lab frame axes. As underlined already, the spin rotation  in the lab frame is convention-independent and
it is precisely the same in the old variables. As a consequence, we should not care to transform the results of the QCD field problem solution  into  the original variables.

The two (covariant and non-covariant) approaches give the same result for real observable effects. The choice between two different approaches is a matter of pragmatics. However, we would like to emphasize a difference in the conceptual background between these two approaches. The non-covariant approach gives additionally a physical insight into the particular laws of nature it deals with. For instance, the dynamical line of arguments that explains the spin rotation in the lab frame is based on the structure of the quantum field equations. In the covariant approach  the dynamics, based on the QCD field equations, is actually hidden in the language of  the  space-time geometry (Minkowski space-time).

\subsubsection{Proper frame view of observations of the lab observer}

Now we wish to continue in our analysis  a little further. We will look for a different way of calculating the spin rotation. We found earlier that the easiest way to derive the BMT equation is to use the Lorentz proper frame. The Wigner rotation of the lab frame axes in the proper frame must be taken into account if we need to know the proper spin dynamics with respect to the lab frame axes. 
In contrast, in the case of the absolute time coordinatization in the proper frame, there is no kinematic Wigner rotation of the lab frame axes with respect to the proper frame axes. The two approaches give, in fact, a different result for spin rotation with respect to the lab frame axes, which must be, however, convention-invariant.  This glaring conflict between results of covariant and non covariant approaches in the proper frame can actually explained. We will see that it is a dynamical line of arguments that explains this paradoxical situation with the relativistic spin rotation.

First we want to rise the following interesting and important point.  The laws of physics in any reference frame should be able to account for all physical phenomena, including the observations made by moving observers. Suppose that an observer in the lab frame performs a spin rotation measurement. To measure the spin direction, a polarimeter at rest in the lab frame can be used.  Suppose we put a charged spinning particle with small $g$ factor through a bending magnet. The lab observer can directly measure the angle of spin rotation at the magnet exit using the polarimeter. The result he observes is consistent with the spin dynamics equation  Eq.(\ref{DDSS9}) \footnote{How does it happen that the construction of the polarimeter never came into discussion before?
In the lab frame the polarimeter is at rest and the field theory involved in the polarimeter operation description is isotropic. We do not need to know any more about the polarimeter operation. In this sense we can discuss in the lab frame  about the spin orientation with respect to the lab frame axes and any detail about the polarimeter  is not needed.}.   
The proper observer sees that polarimeter is moving with a given acceleration and the lab observer, moving with the polarimeter, performs the spin direction measurement. Then, when the polarimeter measurement is analyzed, the proper observer finds that the measured spin rotation angle is nonzero and consistent with    
the spin dynamics equation  Eq.(\ref{DDSS9}), as must be. 

How shell we describe the polarimeter operation  after  Galilean transformations? Suppose that the operation of the polarimeter depends on the electromagnetic field. After the Galilean transformations of the  field equations we obtain  the complicated anisotropic  electrodynamics equations.  The new terms that have to be put into the field equations due to the use of Galilean transformations lead to the same prediction as concerns experimental results: the spin of the particle is rotated with respect to the lab frame axes according to Eq.(\ref{DDSS9}).

In order to predict the result of the  moving polarimeter measurement one does not need to have access to the detailed dynamics of the particle into the polarimeter. 
It is enough to assume  Lorentz covariance of the field theory involved  in the description of the polarimeter operation. 
As before, we use a  mathematical trick for solving the electromagnetic field equations with  anisotropic terms: in order to eliminate these terms in the transformed field equations, we make a change of the variables. Using new  variables  we obtain the phenomenon of spin rotation with respect to the lab frame axes. 

At this point, a reasonable question arises: why in the lab frame view the observed results were analyzed without a description of the polarimeter operation? At first glance the situation is similar to the proper frame setup described above.  The most important difference, however, is that in the lab frame the polarimeter is at rest and the field theory involved in the description of the  polarimeter operation  is isotropic. For instance, when the operation of the polarimeter depends on the electromagnetic field, the field equations are constituted  by the usual Maxwell's equations. In the proper frame we have a similar situation.  Suppose that the polarimeter is at rest with respect to the proper frame and the proper observer performs the spin direction measurement in the special case with $g \to 0$. Then, when the measured data is analyzed, the proper observer finds the trivial result that  there is no spin rotation with respect to the proper frame axes.

\subsection{Incorrect expression for Wigner rotation. Myth about the experimental test}

\subsubsection{Terminology. Thomas precession: correct and incorrect solutions}

Many physicists find that even the terminology is a barrier in the further understanding the properties of Lorentz boosts.  
The composition of two boosts of different planes does not yields boosts, but the product of a boost by a spatial rotation, the latter being known as Wigner rotation. Wigner rotation is sometimes called Thomas rotation (see e.g. \cite{M,JACK}). 
The expression for the Wigner rotation in the lab frame
obtained by  authors of textbooks is given by  $\vec{\delta \Phi} =  (1 - \gamma) \vec{v}\times d\vec{v}/v^2 =	
(1 - \gamma) \vec{\delta \theta}$, which often presented as  $\vec{\omega}_{Th} = d\vec{\Phi}/dt =  
(1 - \gamma) \vec{v}\times d\vec{v}/dt/v^2$ \cite{M}. In other words, the proper frame coordinate performs a precession relative to the lab frame with the velocity of precession  $\vec{\omega}_{Th}$, where $d\vec{v}/dt$ is the acceleration of the spinning particle in the lab frame. This precession phenomenon is called  Thomas precession \cite{M}. From the viewpoint of the generally accepted terminology, Thomas precession is actually a manifestation of the Wigner (Thomas) rotation.
According to expression for Thomas precession in the lab frame presented in textbooks,  the  comoving frame precesses in the opposite direction  with respect to  the  direction of the angular  velocity of the precession $\vec{\omega}_0 =  \vec{v}\times d\vec{v}/dt/v^2$ and $ \omega_{Th} \to -\infty$ in the limit $\gamma \longrightarrow \infty$. 
The theory of relativity shows us that the textbook expression for the Thomas precession  in the lab frame and correct result $\vec{\omega}_{Th} = (1 - 1/\gamma) \vec{v}\times d\vec{v}/dt/v^2$
actually differ both in sign and magnitude.

\subsubsection{Incorrect interpretation of the correct BMT result}

The existence of the usual incorrect expression for the Thomas precession in the lab frame  
has led to incorrect interpretation of the BMT result and, in particular, of the spin dynamics equation  Eq.(\ref{DDSS8}).
Using the incorrect result in \cite{M} for the Thomas precession, 
the  BMT result for a small $g$ factor, Eq.(\ref{DDSS8}), is usually presented as

\begin{eqnarray}
&& \frac{d\vec{s}}{dt} 
= - \left[\left( \gamma - 1\right) \vec{\omega_0} \right]\times\vec{s} = \vec{\omega}_{Th} \times\vec{s} ~ ,\label{DDSS10}
\end{eqnarray}

Note that the results in  Bargmann-Michel-Telegdi  paper \cite{BMT} were obtained by the method of  relativistic "generalization" of the equation of motion for a particle angular momentum in its rest frame.
The Wigner rotation was not considered  at all in \cite{BMT}, because this method allows obtaining the solution for the total particle's spin motion without splitting it into  Larmor and Wigner parts. The interpretation of  Eq.(\ref{DDSS8}) as the Thomas precession  Eq.(\ref{DDSS10}) is presented in textbooks  as alternative approach to the already developed BMT theory.

Frequently, the first stumbling blocks  in the process of understanding and accepting  the correct Wigner (Thomas) rotation theory is a widespread belief that 
the experimental test of the BMT equation is a direct test of the incorrect expression for Thomas precession.
There are many physicists who have already received knowledge about the Thomas precession from well-known textbooks   
and who would say, 
"The extremely precise measurements of the magnetic-moment anomaly of the electron made on highly relativistic electrons are based on the BMT equation, of which the Thomas precession is an integral part, and can be taken as experimental confirmations  of the standard expression for the Thomas precession." 
This misconception about experimental test of the incorrect expression for the Thomas precession in the lab frame is common and pernicious.

The interpretation of the experimentally confirmed equation for the relativistic spin dynamics in the lab frame,  Eq.(\ref{DDSS8}), as the Thomas precession  Eq.(\ref{DDSS10}) is evidently wrong. 
The agreement between BMT result Eq.(\ref{DDSS8}) and Thomas precession results presented in the textbooks is by accident.
Authors of textbooks got the correct BMT result by  using the incorrect expression for the Thomas precession and an incorrect physical argument. This wrong argument is an assumption about  the splitting of spin precession in the lab frame into  dynamics (Larmor) and kinematics (Thomas) parts. In this section we  demonstrated that this splitting can not be obtained in the lab frame.   
It is possible to perform this splitting  only in the Lorentz proper frame where the spinning particle is at rest and the Lorentz lab reference frame  moves with velocity $-\vec{v}$ and acceleration $-\gamma d\vec{v}/d\tau$ with respect to the proper frame axes.

\section{Discussion}

\subsection{Difference between covariant and non-covariant particle trajectories}

\subsubsection{Why use the absolute time coordinatization in accelerator physics?}

Today the statement about correctness of  Galilean transformations  is a "shocking heresy", which offends the "relativistic" intuition and the generally accepted way of looking at special relativity of most physicists. The established way of looking at special relativity is based on Einstein postulates: the principle of relativity and the constancy of the velocity of light.

In the most general space-time geometric  approach to the theory of special relativity, the principle of relativity, in contrast to Einstein formulation of the special relativity, is only a consequence of the geometry of space-time. 
The space-time geometric approach to special relativity deals with all possible choices of coordinates of the chosen reference frames, and therefore the second Einstein postulate, referred to as the constancy of the coordinate speed of light, does not have a place in this  more general formulation. Only in Lorentz coordinates, when Einstein's synchronization of distant clocks and Cartesian space coordinates are used, the coordinate speed of light is isotropic and constant. Thus, the basic elements of the space-time geometric formulation of the special relativity  and the usual Einstein's formulation, are quite different.

The study of relativistic particle motion in a constant magnetic field according to usual accelerator engineering, is intimately connected with the old (Newtonian) kinematics: the Galilean vectorial law of addition of velocities is actually used. A non-covariant  approach to relativistic particle dynamics is based on the absolute time coordinatization, but this is actually a hidden coordinatization.
The absolute time synchronization convention is self-evident and this is the reason why this subject is not discussed in accelerator physics. 
There is a reason to prefer the non-covariant way within the framework of dynamics only. In this approach we have  no mixture of positions and time. This (3+1) dimensional non-covariant particle tracking method is simple, self-evident, and adequate to the laboratory reality.  We demonstrated that there is no principle difficulty with the non-covariant approach in mechanics and electrodynamics. It is perfectly satisfactory. It does not matter which transformation is used to describe the same reality. 
What matter is that, once fixed, such convention should be applied and kept in a consistent way for both dynamics and electrodynamics.

\subsubsection{Where does the old kinematics come from?}

It is general believed that the  integration from initial conditions of the corrected (by introducing a correction factor to the mass)  Newton equation gives the covariant  particle trajectory. However, within the lab frame the motion of particles in a constant magnetic field looks precisely  the same as predicted  by Newtonian kinematics: relativistic kinematics effects do not have a place in this description. In conventional particle tracking  a particle trajectory  $\vec{x}(t)$ can be seen from the lab frame as the result of successive Galileo boosts that track the motion of the accelerated  particle. 
Therefore, the trajectory $\vec{x}(t)$  cannot be identified with the covariant trajectory $\vec{x}_{cov}(t)$ even if, at first glance, it appears to be derived following the covariant prescription. 

Where does the old kinematics come from? In our previous publication \cite{OURS6} we examined the question about where the hidden assumption of absolute time coordinatization was introduced. We found that
the commonly accepted derivation of a version of Newton's second law  corrected by the relativistic factor $\gamma = 1/\sqrt{1 - v^2/c^2}$   from the covariant equation has one delicate point. 
The corrected Newton equation is not derived from the covariant equation only. 
In this equation, the restriction  $\vec{p}  = m\vec{v}/\sqrt{1 - v^2/c^2}$ has already been imposed. Why should there be this restriction? 
We know that in accepted covariant approach, the solution of the dynamics problem for the momentum in the lab frame makes no reference to the three-dimensional velocity. In fact, the manifestly covariant (four-tensor)  equation of motion for a relativistic charged particle under the action of the four-force tells us that the force is the rate of change of the momentum $\vec{p}$, but does  not tell us how momentum varies with speed. 
The four-velocity cannot be decomposed into  $u = (c\gamma, \vec{v}\gamma)$  when we deal with a particle accelerating along a curved trajectory in the Lorentz lab frame. 

The components of the momentum four-vector $p_\mu = (E/c,\vec{p})$ behave, under transformations from one Lorentz frame to another, exactly in the same manner as the components of the four-vector event $x = (x_0,\vec{x})$.  Surprises can surely be expected when we return from the four-vectors language  to the three-dimensional velocity vector $\vec{v}$, which can be represented in terms of the components of four-vector as $\vec{v}/c = d\vec{x}/dx_0$.  
In contrast with the pseudo-Euclidean four-velocity space, the relativistic three-velocity space is a three-dimensional space with constant negative curvature, i.e. three-dimensional space with Lobachevsky geometry. 
 
Then, where does the decomposition  $u = (c\gamma, \vec{v}\gamma)$ come from? We see that what we have to do is to assume that $d\tau = dt/\gamma$. In fact,  to do that we have to find  $u_\mu = dx_\mu/d\tau = \gamma dx_\mu/dt = (c\gamma, \vec{v}\gamma)$.
The presentation of the time component simply as  the relation $d\tau = dt/\gamma$ between proper time and coordinate time is based on the hidden assumption that the type of clock synchronization, which provides the time coordinate $t$ in the lab frame, is based on the use of the absolute time convention.

\subsubsection{Incorrect expansion of the relation  $d\tau = dt/\gamma$  to an arbitrary motion}

Authors of textbooks are dramatically mistaken in their belief about the usual momentum-velocity relation.
From the theory of relativity follows that the equation $\vec{p}_{cov}  = m\vec{v}_{cov}/\sqrt{1 - v^2_{cov}/c^2}$  does not hold for a curved trajectory in the Lorentz lab frame.
Many experts who learned the theory of relativity using  textbooks will find this statement disturbing at first sight.

It is  well known that for rectilinear accelerated motion the usual momentum-velocity relation holds. 
In fact, for the rectilinear motion the combination of the usual momentum-velocity relation and the covariant three-velocity transformation (according to Einstein's law of velocity addition) is consistent with the covariant three-momentum transformation and both  (non-covariant and covariant) approaches produce the same trajectory.

We can see why by examine the transformation of the three velocity in the theory of relativity. For a rectilinear motion, this transformation is performed as
$v = (v'+V)/(1 + v'V/c^2)$. The "summation" of two velocities is not just the algebraic sum of two velocities, but it is "corrected" by $(1 + v'V/c^2)$. 
The relativistic factor $1/\sqrt{1 - v^2/c^2}$ is given by:
$1/\sqrt{1 - v^2/c^2} = (1 + v'V/c^2)/(\sqrt{1 - v'^2/c^2}\sqrt{1 - V^2/c^2})$. The new momentum is  then simply $mv$ times the above expression. But we want to express the new momentum in terms of the primed momentum and energy, and we note that $p = (p' + \mathcal{E}'V/c^2)/\sqrt{1 - V^2/c^2}$. Thus, for a rectilinear motion, the combination of  Einstein addition law for parallel velocities and the usual momentum-velocity relation is consistent with the covariant momentum transformation.

This result was incorrectly extended to an arbitrary trajectory. 
Like it happens with  the composition of Galilean boosts, collinear Lorentz boosts commute.
This means that the resultant of successive collinear Lorentz boosts is independent of the transformation order. On the contrary, Lorentz boosts in different directions do not commute. 
As discussed above, the composition of non-collinear boosts is equivalent to a boost followed  by a spatial rotation, the Wigner rotation. The  Wigner rotation is relativistic effect that does not have  a non-covariant analogue. One of the consequences of non-commutativity of non-collinear Lorentz boosts is the unusual momentum-velocity relation
$\vec{p}_{cov}  \neq m\vec{v}_{cov}/\sqrt{1 - v^2_{cov}/c^2}$, which also does not have any non-covariant analogue. 

The theory of relativity shows that the unusual momentum-velocity relation discussed above is related  with the acceleration along curved trajectories. In this case there is a difference between  covariant and non-covariant  particle trajectories. One can see that this essential point has never received attention by the physical community. Only the solution of the dynamics equations in covariant form gives the correct coupling between the usual Maxwell's equations and particle trajectories in the lab frame. 
A closer analysis of the concept of velocity, i.e. a discussion of the methods by which a time coordinate can actually be assigned in the lab frame, opens up the possibility of a description of such physical phenomena as radiation from a relativistic electron accelerating along a curved trajectory in accordance with the theory of relativity.

\subsection{Why did the error in radiation theory remain so long undetected?}

Accelerator physicists, who try to understand the situation related to the use of the theory of relativity in the synchrotron radiation phenomena, are often troubled by the fact that the difference between covariant and non-covariant particle trajectories was never understood, and that nobody realized that  there was a contribution to the synchrotron radiation from relativistic kinematics effects. Accelerator physics was always thought  in terms of the old (Newtonian)  kinematics that is not compatible with Maxwell's equations.  At this point, a reasonable question arises: since storage rings are designed without accounting for the relativistic kinematics effects, how can they actually operate?  In fact, electron dynamics in storage ring is greatly influenced by the emission of radiation. We have already answered this question in great detail in our previous publication \cite{OURS6}.

\subsubsection{Covariant particle tracking. Great simplification in ultrarelativistic limit}

For an arbitrary parameter $v/c$ covariant calculations of the radiation process are very difficult. There are, however, circumstances in which  calculations can be greatly simplified. An example of such circumstance is a synchrotron radiation setup.
Similar to the non-relativistic asymptote, the ultrarelativistic asymptote is also  characterized by the essential simplicity of the covariant calculation. The reason is that the ultrarelativistic assumption implies the paraxial approximation. Since the formation length  of the radiation is much longer than the wavelength, the radiation is emitted at small angles of order $1/\gamma$ or even smaller, and we can therefore enforce the small angle approximation. We assume that the transverse velocity  is small compared to the velocity of light. In other words, we use a second order relativistic approximation for the transverse motion.
Instead of small (total) velocity parameter $(v/c)$ in the non-relativistic case, we use a small transverse velocity parameter  $(v_{\perp}/c)$.
The next step is to analyze the longitudinal motion, following the same method. We should remark that the analysis of the longitudinal motion in a synchrotron radiation setup is very simple. If we evaluate the transformations up to  second order  $(v_{\perp}/c)^2$, the relativistic correction in the longitudinal motion does not appear. 

According to the covariant approach, the various relativistic kinematics effects concerning to the synchrotron radiation setup turn up in successive orders of approximation.

In the first order $(v_{\perp}/c)$. - relativity of simultaneity. Wigner rotation, which in the ultrarelativistic approximation appears in the first order already, and results directly from the relativity of simultaneity.

In the second order $(v_{\perp}/c)^2$. - time dilation. Relativistic correction in the law of composition of velocities, which already appears in the second order, and results directly from the time dilation.

The first order kinematics term $(v_{\perp}/c)$ plays an essential role only in the description of extended (macroscopic) relativistic objects. But up to the 21 st century there were no macroscopic objects possessing relativistic velocities.
An XFEL is an example where improvements in accelerator technology makes it possible to develop ultrarelativistic macroscopic objects with an internal structure (modulated electron bunches), and the first order kinematics term  $(v_{\perp}/c)$ plays an essential role in their description. 
We demonstrated that relativistic kinematics enters XFEL physics in a most fundamental way through the Wigner rotation of the modulation wavefront, which,  in ultrarelativistic approximation, is closely associated to the relativity of simultaneity.

The first order kinematics term $(v_{\perp}/c)$ plays an essential role only in the description of the coherent radiation from a modulated electron beam. In a storage ring the distribution of the longitudinal position of the electrons in a bunch is essentially uncorrelated. In this case, the radiated fields due to different electrons are also uncorrelated and the average power radiated is a simple sum of the radiated power from individual electrons; that is we sum intensities, not fields. 
A motion of the single ultrarelativistic electron in a constant magnetic field, according to the theory of relativity, influences the kinematics terms of the second order $(v_{\perp}/c)^2$ only.

\subsubsection{ Covariant ultrarelativistic electron tracking in a constant magnetic field} 

In \cite{OURS6} we attempted to answer the question of  why the error in radiation theory should have so long remained undetected. 
We have seen that due to  a combination of the ultrarelativistic (i.e. paraxial) approximation  and a very special (cylindrical) symmetry of the bending magnet setup there is a  beautiful cancellation of the relativistic kinematics effects. That means that the synchrotron radiation from  bending magnets does not show any sensitivity to the difference  between covariant and non-covariant particle trajectories. 
But in the 21st century with the operation XFELs  this situation changes.
An XFEL is an example where the first order kinematics term $(v_{\perp}/c)$ plays an essential role in the description of the coherent radiation from the modulated electron beam  and,  in this case, covariant coupling of fields and particles predicts  an effect in complete contrast to the conventional treatment.

We want to emphasis an important features of our covariant analysis of the radiation process in a constant magnetic field:  for a semi-relativistic parameter $v/c$ the  cancellation of the relativistic kinematics effects does not take place 
and the conventional coupling of Maxwell's electrodynamics and non-covariant particle trajectory is absolutely incapable of correctly describing semi-relativistic particle radiation in a constant magnetic field. There is a difficulty with such range of parameters in the covariant theory. It is clear that, without paraxial approximation, things get very complicated and
the covariant result would be rather difficult to calculate.

In order to  understand the origin of the cancellation  without much calculations, let us  
start by considering an ultrarelativistic electron moving along a circular trajectory that lies in the $(x,z)$-plane and  tangent to the $z$ axis. 
Note that the geometry of the electron motion has a cylindrical symmetry, with the vertical axis going through the center of the circular orbit. The observer is assumed to be located in the vertical plane tangent to the circular trajectory at the origin. 
Because of cylindrical symmetry, in order to calculate spectral and angular photon distributions, it is not necessary to consider an observer in the lab frame at a generic location. 

In conventional particle tracking  a particle trajectory  $\vec{x}(t)$ can be seen from the lab frame as the result of successive Galileo boosts that track the motion of the accelerated  particle in a constant magnetic field. The usual Galileo rule for addition of velocities is used to determine the Galileo boosts tracking a particular particle, instant after instant, through its motion along the curved trajectory.
In section 3.3 we described  a general method for finding solution to the electrodynamics problem in the case of the absolute time coordinatization. Since the Galilean transformation $x = x' + vt', ~ t = t'$, completed by the introduction of the new variables $ct_n = \left[ \sqrt{1-v^2/c^2}ct +  (v/c)x/\sqrt{1-v^2/c^2}\right]$, and
$x_n = x/\sqrt{1 - v^2/c^2}$, is mathematically equivalent to a Lorentz transformation
$x_n = \gamma(x' + vt')~ , t_n = \gamma(t' + vx'/c^2)$, it obviously follows that transforming to new variables $x_n, t_n$ leads to the usual Maxwell's equations.

In ultrarelativistic (paraxial) approximation we evaluate transformations working only up to the order of $v^2_x/c^2$. The restriction to this order provides an essential simplicity of calculations. 
We can interpret a manipulation of the rule-clock structure in the lab frame simply as a change of variables according to the transformation $x_d = \gamma_x x$, $t_d = t/\gamma_x + \gamma_x xv_x/c^2$, where we are dealing with a second order approximation and $\gamma_x = 1 + v_x^2/(2c^2)$.
The overall combination of Galilean transformation and variable changes actually yields to the  transverse Lorentz transformation (see section 3.3 for more detail).

In order to keep Lorentz coordinates in the lab frame, as discussed before, we need only to perform a clock resynchronization by introducing the time shift $\Delta t = t_d - t =  - [v_x^2/(2c^2)]t + xv_x/c^2$. The relativistic correction to the particle's offset "$x$" does not appear in this expansion order, but only in the order of $v_x^3/c^3$ and $x_d = x$ in our case of interest. 
To finish our analysis we only need to find a relativistic correction to the longitudinal motion. We remark again that if we evaluate the transformations up to the second order $(v_x/c)^2$, the relativistic correction in the longitudinal motion does not appear.

In  the ultrarelativistic approximation, we have a uniform acceleration of the electron $a = v^2/R = c^2/R$ in the transverse direction. We can, then, write velocity and offset of the electron as  $v_x = at$, $x = at^2/2$. We have now all quantities we need. Let us substitute them all together in the relativistic time shift:  $\Delta t =  t_d - t = - a^2t^3/(2c^2) + a^2t^3/(2c^2) = 0$, meaning that there is no time shift! We do not need to use the covariant particle tracking for describing  bending magnet radiation. Why should that be? Usually, such a beautiful cancellation is found to stem from a deep underlying principle. Nevertheless, in this case there does not appear to be any such profound implication.  
This is a coincidence. It is because we have deal with uniform acceleration in the transverse direction using a second order (paraxial) approximation when an electron is moving along an arc of a circle.

We consider now some physical situation in which the second order kinematics term $v^2_x/c^2$ plays an essential role in the description of the spontaneous synchrotron radiation. 
Let us discuss the bending magnet radiation from a single electron with a kick with respect to the nominal orbit in $(x,z)$-plane. In this case, we additionally have a translation along the $y$-axis with constant velocity $v_y = v\theta_k$. We can, then, write the offset of the electron as  $y = \theta_k z$. Let us substitute velocity and offset in the relativistic time shift: $\Delta t = t_d - t = -\theta_k^2z/(2c)
+ \theta_k^2z/c = \theta_k^2z/(2c)$.  In this case  there is a difference which leads to the red shift of the critical frequency of the synchrotron radiation. Synchrotron radiation from bending magnets is emitted in a broad spectrum, and its angular-spectral density distributions are not sensitive to red shift of the critical frequency. The possibility of using narrow bandwidth sources in experimental study of the red shift in synchrotron radiation spectrum looks quite attractive. 

In  \cite{OURS6} we also discussed the undulator radiation.
According to the correct coupling of fields and particles, there is a remarkable prediction of synchrotron radiation theory concerning the  emission of undulator radiation from a single electron with  and without kick. 
Namely, when a kick is introduced, there is a red shift in the resonance frequency of the undulator radiation in the velocity direction.

\subsubsection{Results of experiment}


The fact that our theory  predicts reality in a satisfactory way is well-illustrated by comparing the prediction we  made in \cite{OURS6} with the results of an experiment involving "X-ray beam splitting" of a circularly-polarized XFEL pulse from the linearly-polarized XFEL background pulse, a technique used in order to maximize the degree of circular polarization at XFELs \cite{NUHN}.
It apparently demonstrated that after a modulated electron beam is kicked on a large angle compared to the divergence of the XFEL radiation, the modulation wavefront is readjusted along the new direction of the motion of the kicked beam. Therefore, coherent radiation from the undulator placed after the kicker is emitted along the kicked direction practically without suppression. 

In the framework of the conventional theory, there is also a second outstanding puzzle concerning the beam splitting experiment at the LCLS. In accordance with conventional undulator radiation theory, if the modulated electron beam is at perfect  resonance without kick, then after the kick the same modulated beam must be at perfect resonance in the velocity direction. 
However, experimental results clearly show that when the kick is introduced
there is a red shift in the resonance wavelength. The maximum power of the coherent radiation is reached when the undulator is detuned to be resonant to the lower longitudinal velocity after the kick \cite{NUHN}.    

It should be remarked that any linear superposition of given radiation fields from single electrons conserves single-particle characteristics like parametric dependence on undulator parameters and polarization. Consider a modulated electron beam kicked by a weak dipole field before entering a downstream undulator. Radiation fields generated by this beam can be seen as a linear superposition of fields from individual electrons. Now experimental results clearly show that there is a red shift in the resonance wavelength for coherent undulator radiation when the kick is introduced. It follows that the undulator radiation from a  single electron is red shifted when the kick is introduced. This argument suggests that results of the beam splitting experiment in reference \cite{NUHN} confirm our  correction for spontaneous undulator emission.
In fact, one of the immediate consequences of our theory is the occurrence of the non-zero red shift of the resonance wavelength when the kick angle has nonzero value
\footnote{It has been claimed in a recent paper \cite{MA} that accounting for the quadrupole lattice in the baseline XFEL undulator  it is possible to obtain a mechanism for the modulation wavefront to tilt forward, towards the new direction of propagation. 
A theoretical analysis of an XFEL driven by an electron beam with wavefront tilt was presented in \cite{MA}, based on the use the usual Maxwell's equations. In fact, the Maxwell solver was used as a part of  the standard simulation code.
We state that this approach is fundamentally incorrect. In ultrarelativistic asymptote a modulation wavefront tilt is absurd with the viewpoint of Maxwell's electrodynamics. 
In the  case of an XFEL we deal  with an ultrarelativistic electron beam  and within the Lorentz lab frame (i.e. within the validity of the Maxwell's equations) the tilted modulation wavefront is at odds with the principle of relativity.}.

\subsection{First  practical applications of the Wigner rotation theory}

\subsubsection{BMT equation}

Let us now review the subjects discussed in the past few sections.
We considered the widespread misconception that if a particle with spin has no magnetic moment ($g \to 0$), the spin rotation in the lab frame is purely kinematic in nature. 
To quote e.g. Anderson \cite{A}: " ... if $F^{\mu\nu} \neq 0$, even with $g = 0$, we see that $dS^{\mu}/d\tau \neq 0$. Thus a spinning charge particle will precess  in an electromagnetic field 
even if it has no magnetic moment. This precession is a pure relativistic effect known as the Thomas precession."

We discussed how authors of textbooks got the correct BMT result by the incorrect expression for the Thomas precession and an incorrect physical argument. This wrong argument is an assumption about the  splitting of the spin precession in the lab frame into dynamics (Larmor) and kinematics (Thomas) parts. This splitting can not be realized in the lab frame for the following reason: the starting point of the BMT theory is the phenomenological dynamics law 
$d\vec{s}/d\tau =  eg\vec{s}\times \vec{H}_R/(2mc)$, 
which is the equation of motion for the angular momentum in its rest frame (i.e. with respect to the rest frame axes). It is phenomenological because the microscopic interpretation of the (anomalous) magnetic moment  of a particle is not given. The BMT equation is a relativistic generalization of the phenomenological dynamics law. It is valid for any given Lorentz  frame, for example for  the Lorentz lab frame.  In the lab frame, the  BMT equation is a phenomenological dynamics equation for the spin motion with  respect to the lab frame axis even at $g \to 0$.

Another argument for the dynamics origin of the spin rotation in the lab frame at $g \to 0$
is that the relativistic kinematic effects are coordinate (i.e. convention-dependent) effects and have no exact objective meaning. In the case of  Lorentz coordinatization, one will experience e.g. the Thomas precession phenomenon. In contrast to this, in the case of  absolute time coordinatization there are no relativistic kinematics effects, and therefore no Thomas precession will be found. However, the spin orientation at $g \to 0$ with respect to the lab frame axes at each point of the particle path in the lab frame  has exact objective meaning (i.e. it has no kinematical origin).

With the wording "spin orientation with respect to the lab frame axes in the lab frame" we mean that the lab observer can directly measure the spin orientation with respect to the lab frame axes using a polarimeter. In the lab frame, the polarimeter is at rest and the field theory involved in the polarimeter operation description is isotropic. We do not need to know any more about the polarimeter operation. It is in this sense, we can discuss in the lab frame  about the spin orientation with respect to the lab frame axes as physical reality.

In contrast, the spin orientation with respect to the lab frame axes in the proper frame ( and also  the spin orientation with respect to the proper frame axes in the lab frame) has no direct physical meaning. In fact, if we use the Lorentz coordinatization in the proper frame there is a Thomas precession of the lab frame axes in the proper frame.
In the case of  absolute time coordinatization in the proper frame, there is no kinematic Thomas precession of the lab frame axes with respect to the proper frame axes. The two coordinatizations give, in fact, a different result for spin rotation with respect to the lab frame axes. 

At this point a reasonable question arises: why in the lab frame the spin orientation with respect to the lab frame axes has physical meaning, but the same orientation in the proper frame does not? The answer is that in the lab frame an observer who performs spin orientation measurement is at rest with respect to the lab reference frame. This situation is symmetrical with respect to a change of the reference frames. In fact, the spin orientation measurement with respect to the proper frame axes in the proper frame has exact objective meaning (i.e. it has dynamical origin) and we observe the same result no matter how the lab frame rotates with respect to the proper frame.

The argument that the result of  spin orientation measurements with respect to the lab frame axes in the proper frame is paradoxical runs  something like this: 
the laws of physics in any one reference frame should be able to account for all physical phenomena, including the observations made by moving observers. Suppose that an observer in the lab frame performs a spin rotation measurement. Viewed from the proper frame, the two proper frame coordinatizations give a different result for the spin rotation with respect to the lab frame axes in the lab frame which must be convention-invariant. 

Nature doesn't see a paradox, however, because the proper observer sees that the lab polarimeter is moving on an accelerated motion, and the lab observer, moving with the polarimeter, performs the spin direction measurement. In order to predict the result of the  moving polarimeter measurement one does not need to have access to the detailed dynamics of the particle into the polarimeter. 
It is enough to assume the Lorentz covariance of the field theory involved  in the description of the polarimeter operation.

In the Lorentz proper frame the field theory involved in the description of the (lab) polarimeter operation is isotropic. Clearly, in the case of Lorentz coordinatization we can discuss in the proper  frame  about the spin orientation with respect to the lab frame axes as a prediction of the measurement made by the lab observer.   

Now the question is, what is the prediction of the proper observer in the case of absolute time coordinatization? How shell we describe the polarimeter operation  after the Galilean transformations? After the Galilean transformations  we obtain  the complicated (anisotropic ) field equations.  
The new terms that have to be put into the field equations due to the use of Galilean transformations lead to the same prediction as concerns experimental results: the spin of the particle is rotated with respect to the lab frame axes according to the Lorentz coordinatization prediction. Let us examine in a little more detail how this spin rotation comes about. As usual, in the case of absolute time coordinatization, we are  going to make a  mathematical trick for solving  the field  equations with anisotropic terms. In order to eliminate these terms  we make a change of the variables. Using new  variables  we obtain the  phenomenon of spin rotation with respect to the lab frame axes in the lab frame.

\subsubsection{Modulation wavefront rotation}

There is common misconception that the modulation wavefront orientation has objective meaning.
Let us consider the predictions of the  conventional XFEL  theory in the case of non-collinear electron beam motion. As well-known, upon conventional particle tracking, after an electron beam is kicked by a weak dipole magnet, the kick results in a difference between the directions of the electron motion and the normal to the modulation wavefront i.e. in a wavefront tilt. 

When the trajectories of the particles calculated in a Lorentz reference frame,  they must include such relativistic kinematics effect as relativity of simultaneity. In the ultrarelativistic asymptote, the orientation of the modulation wavefront , i.e the orientation of the plane of simultaneity, is always perpendicular to the electron beam velocity when the evolution of the modulated electron beam is treated using Lorentz coordinatization. 

The tilt of the modulation wavefront is not a real observable effect. Indeed, if we  couple our particle system with electromagnetic fields in accordance with the principle of relativity, we find that  coherent undulator radiation from the modulated electron beam is always emitted in the kicked direction, independently of the coordinatization.

It is not difficult to see this using a Lorentz coordinate system where Maxwell's equations are valid and the modulation wavefront is always perpendicular to the beam velocity. In Maxwell's electrodynamics, coherent radiation is always emitted in the direction of the normal to the modulation wavefront. Indeed, we may consider the amplitude of the beam radiated as a whole to be the resultant of radiated spherical waves. This is because Maxwell's theory has no intrinsic anisotropy.

We can derive the same results for the direction of radiation propagation  with the help of Galilean transformations. According to this old kinematics, the orientation of the modulation wavefront is unvaried. However, Maxwell's equations do not remain invariant with respect to Galilean transformations, and the choice of the old kinematics implies the use of anisotropic field equations. In this case the secondary waves (wavelets) are not spherical. As a result, the wavefront remains plane but the direction of propagation is not perpendicular to the wavefront. In other words, the radiation beam motion and the radiation wavefront (phase front) normal have different directions.

In this study we  presented the two practical applications of the Wigner rotation theory in the lab frame ( XFEL theory) and in the proper frame (BMT theory).
A close look at the physics of these two subjects shows things which are common to these phenomena:

1. Orientation of the lab frame axes in the proper frame (orientation of the modulation wavefront in the lab frame) is not real observable. 

2. The final calculations of the spin orientation in the lab frame (the direction of the radiation propagation) does not depends on the synchronization convention.

3. The final result of calculations  does not change, but the orientation of the lab frame axes with respect to the proper frame axes (the wavefront orientation in the lab frame)   does, depending on the choice of coordinatization.

If we look more closely at the physics, we would see aspects that are not common to these phenomena. The equation we found for the spin motion is only a generalization of a phenomenological law, and the details of underlaying complexities related with a microscopic  model are hidden in the BMT equation. That is the usefulness  of the principle of relativity - it permits us to make predictions, even about things that otherwise we do not know much about. In contrast, we know why the modulated electron beam coherently radiate in an XFEL and what its machinery is. In this case we present results of first-principles calculations.

\subsection{On the advanced "paradox" related to the coupling fields and particles}

\subsubsection{Setup description}

We have already discussed bending magnet radiation from an ultrarelativistic electron with angular deflection with respect to the nominal orbit.
We  now want to point out that
there are two different sets of initial conditions resulting in  the same uniform translation along the magnetic field direction in the Lorentz lab  frame. 
Suppose that an electron moves, initially, at  ultrarelativistic velocity $v$ parallel to the $z$- axis upstream a uniform magnetic field (i.e. bending magnet) directed along the $y$-axis. In other words, we start by considering  an electron moving along a circular trajectory that lies in the $(x,z)$-plane.  We then rotate the magnetic field vector $\vec{H}$ in the $(y,z)$-plane by an angle $\theta$, assuming that rotation angle is small. We consider a situation in which the electron is in uniform motion  with velocity $v\theta$ along the magnetic field direction. It is clear that if we consider the radiation from an electron moving on a circular orbit, the introduction of the magnetic field vector rotation will leave the radiation properties unchanged \footnote{This is plausible, if one keeps in mind that after rotating the bending magnet, the electron has the same velocity and emits radiation in the velocity direction owing to the Doppler effect. After the rotation,  correction to the curvature radius $R$ is only of order $\theta^2$ and can be neglected. }.

Now we consider another situation. Let us see what happens with a kicker, which is installed in the straight section upstream of the bending magnet and is characterized by a small kick angle $(\gamma\theta)^2 \ll 1$. 
When the kick in the $y$ direction is introduced, there is a red shift of the critical wavelength which arise because, according to Einstein's addition velocities law, the electron velocity decreases from $v$ to $v - v\theta^2/2$ after the kick \footnote{It is easy to see that the acceleration in the kicker field yields an electron velocity increment $v_x = v\theta$ parallel to the $x$-axis and $\Delta v_z = - v\theta^2/2$ parallel to the $z$-axis. If we neglect terms in $(\gamma v_x/c)^3 = (\gamma\theta)^3$, the relativistic correction in the composition of increments does not appear in this approximation (i.e. these two Lorentz boosts commute). It is well known that a Lorentz boost with non relativistic velocity $v_x$ leads simply to a rotation of the particle velocity $v_z = v - v\theta^2/2$ of the angle $\theta = v_x/c$.}    
The red shift of the critical frequency $\omega_c$ can be expressed by the formula $\Delta\omega_c/\omega_c = - (3/2)\gamma^2\theta^2$. We see a second order correction $\theta^2$ that is, however, multiplied by a large factor $\gamma^2$.   
The result of the covariant approach clearly depends on the absolute value of the kick angle $\theta$ and the radiation along the velocity direction has a red shift only when the kick angle has nonzero value.

The difference between these two situations, ending with a final uniform translation along the direction of the magnetic field is very interesting.
It comes about as the result of the difference between two Lorentz coordinate systems in the lab frame. By trying to accelerate the electron upstream the bending magnet we have changed Lorentz coordinates for that particular source.
We know that in order to keep a Lorentz coordinates system in the lab frame after the kick we need to perform a clock resynchronization. So we should expect the electron velocity to be changed.
The difference between the two setups 
is understandable. When we do not perturb the electron motion upstream of the bending magnet, no clock resynchronization takes place, while when we do perturb the motion, clock resynchronization is introduced. We must conclude that 
when we accelerate the electron in the lab frame upstream the bending magnet, the information about this acceleration is included into the covariant trajectory.

It should be note, however, that there is another satisfactory way of explaining the red shift.  
We can reinterpret this result with the help of a non-covariant treatment, which deals with non- covariant particle trajectories, and with Galilean transformations of the electromagnetic field equations. According to non-covariant particle tracking, the electron velocity is unvaried. However, Maxwell's equations do not remain invariant with respect to Galilean transformation,  and the velocity of light has increased from $c$, without kick, to $c(1 + \theta^2/2)$ with kick.  The reason for the velocity of light being different from the electrodynamics constant $c$ is due to the fact that, according to the absolute time convention, the clocks after the kick are not resynchronized.
Now everything fits together, and our calculations  show that covariant and non-covariant treatments (at the correct coupling fields and particles) give the same result for the red shift prediction, which is obviously convention-invariant and  has direct objective meaning.

\subsubsection{A "paradox"}

We would now like to describe an apparent paradox. A paradox is a statement that is seemingly contradictory, but, in reality, expresses truth without contradiction. 
When the situation is described as we have done it here, there doesn't seem to be any paradox at all. The argument that the difference between these two situations, ending with a final uniform translation along the magnetic fields direction, 
is paradoxical can be summarized  in the following way: in the case of absolute time coordinatization in the lab frame,  the initial conditions at the bending magnet entrance are apparently identical. In fact, the magnitude of the electron velocity and the orientation of the velocity vector with respect to the magnetic field vector  are identical in both setups. We must conclude that when we accelerate the electron in the lab frame upstream the bending magnet, the information about this acceleration is not included into the non-covariant trajectory. Where is the information about the electron acceleration recorded in the case of  absolute time coordinatization? Since  an electron is a structureless particle, the situation seems  indeed paradoxical.

\subsubsection{Solution to the "paradox"}

Electrodynamics deals with observable quantities. 
Let us consider the measurement of the red shift in the bending magnet radiation from our  kicked electron.
We can measure the  accurate value of the red shift using a spectrometer in the lab frame, and this leads to a  description of the setup in the space-frequency domain.

Suppose we have a uniformly moving electron. 
The fields associated to an  electron with constant velocity exhibit an interesting behavior when the speed of the charge approaches that of light. Namely, in the space-frequency domain there is an equivalence of the fields of a relativistic electron and those of a beam of electromagnetic radiation.  In fact, for a rapidly moving electron we have  nearly equal transverse and mutually perpendicular electric and magnetic fields. These are indistinguishable  from the fields of a beam of radiation. This virtual radiation beam has a macroscopic transverse size of order $\lambdabar\gamma$. An ultrarelativistic electron at synchrotron radiation facilities, emitting  at nanometer-wavelengths (we work in the space-frequency domain) has indeed a macroscopic  transverse size of order of 1 $\mu$m. The field distribution of the virtual radiation beam is described by the Ginzburg-Frank formula.

When the electron passes through a kicker, its fields are perturbed, and now include information about the acceleration. 
According to the old kinematics, the orientation of the virtual radiation phase front is unvaried. However, Maxwell's equations do not remain invariant with respect to Galilean transformations and, as discuss throughout this paper,  the choice of the old kinematics implies using  anisotropic field equations. As a result, the phase front remains plane but the direction of propagation is not perpendicular to the phase front. In other words, the radiation beam motion and the radiation phase front normal have different directions. Then, having chosen the absolute time synchronization, electrodynamics  predicts that the  virtual radiation beam propagates in the kicked direction with the phase front tilt $\theta_k$. 
This is the key to the "paradox" discussed here. The information about the electron acceleration is recorded in the perturbation of the self-electromagnetic fields of the electron. Mathematically information is recorded in the phase front tilt of the virtual radiation beam.

\subsubsection{What does space-time geometry explain?}

It is important to stress at this point that the dynamical line of argument discussed here explains what the Minkowski geometry physically means. The pseudo-Euclidean geometric structure of space-time is only an interpretation of the behavior of the dynamical matter fields in the view of different observers, which is an observable, empirical fact. It should be clear that the relativistic properties of the dynamical matter fields are fundamental, while the geometric structure is not.  Dynamics, based on the field equations, is actually hidden in the language of kinematics. The Lorentz covariance of the equations that govern the fundamental interactions of nature is an empirical  fact, while the postulation of the pseudo-Euclidean geometry of space-time is a mathematical interpretation of it that yields the laws of relativistic kinematics: at a fundamental level this postulate is, however, based on the way  fields behave dynamically.

\section{Conclusions}

In this study we presented, for the first time, two practical applications of the Wigner rotation theory in the lab frame ( XFEL theory) and in the proper frame (BMT theory). Understanding these two applications is a pre-requisite for the understanding of many other phenomena that occur in complex situations relating to the use of the theory of relativity in accelerator physics.

\section{Acknowledgments}

We acknowledge many useful discussions with Gianluca Geloni and Vitaly Kocharyan. We are also indebted to Gianluca Geloni for carefully reading this manuscript, as well as for his continuous advice during its development.

\end{document}